\documentclass[longbibliography,groupedaddress,showpacs,showkeys,amssymb,eqsecnum,aps,nofootinbib]{revtex4}
\usepackage[]{graphicx}
\usepackage[]{graphics}
\usepackage{amsmath}
\usepackage{epsf}                                                                                           
\usepackage{color}                   
\usepackage{verbatim}                                                                         
\usepackage{hyperref}

\newcommand{\be}{\begin{equation}}

\newcommand{\ee}{\end{equation}}

\newcommand{\dd}{\mathop{d^{d}x}}

\begin{document}                                                                                              

\title{On Restricting First Order Form of Gauge Theories to One-Loop Order}

\author{F. T. Brandt}  
\email{fbrandt@usp.br}
\affiliation{Instituto de F\'{\i}sica, Universidade de S\~ao Paulo, S\~ao Paulo, SP 05508-090, Brazil}

\author{J. Frenkel}
\email{jfrenkel@if.usp.br}
\affiliation{Instituto de F\'{\i}sica, Universidade de S\~ao Paulo, S\~ao Paulo, SP 05508-090, Brazil}

\author{S. Martins-Filho}   
\email{sergiomartinsfilho@usp.br}
\affiliation{Instituto de F\'{\i}sica, Universidade de S\~ao Paulo, S\~ao Paulo, SP 05508-090, Brazil}

\author{D. G. C. McKeon}
\email{dgmckeo2@uwo.ca}
\affiliation{
Department of Applied Mathematics, The University of Western Ontario, London, Ontario N6A 5B7, Canada}
\affiliation{Department of Mathematics and Computer Science, Algoma University, 
Sault Ste.~Marie, Ontario P6A 2G4, Canada}

\date{\today}

\begin{abstract}
    The first order form of the Yang-Mills and Einstein-Hilbert
    actions are quantized, and it is shown how Green's functions
    computed using the first and the second order form of these theories
    are related. Next we show how by use of Lagrange multiplier
    fields (LM), radiative effects beyond one-loop order can be
    eliminated. This allows one to compute Green's functions exactly
    without loss of unitarity. The consequences of this restriction on
    radiative effects are examined for the Yang-Mills and
    Einstein-Hilbert actions. In these two gauge theories, we find
    that the quantized theory is both renormalizable and unitary once
    the LM field is used to eliminate effects beyond one-loop order.
\end{abstract}                                   


\maketitle

\section{Introduction}

Quantization of gauge theories using the Faddeev--Popov (FP) procedure in conjunction with the path integral is most often done using the second order form of these theories; this is described in numerous text books \cite{taylor:1976b, weinberg:book1995, peskin_scroeder}. However, in the Yang-Mills (YM) action, this involves having rather complicated three and four point vertices; with the Einstein-Hilbert (EH) action one is faced an infinite series of vertices. 
In both of these gauge theories, by going the first order form one only encounters a single three point vertex, and this vertex is independent of momentum \cite{McKeon:1994ds,Buchbinder:1985jc,Brandt:2015nxa,Brandt:2016eaj}. However, in the first order formalism there are now two propagating fields. In the YM action these are the vector potential and the field strength while in the EH action the metric and affine connection propagate independently.

The equivalence of the first and second order formalism is easily
shown to be true at the classical level by using the equation of
motion to eliminate one of the fields in the first order form of the
action. Once this is done, the second order form of the action is
recovered. However, establishing how the quantized version of the
first and the
second order form of gauge theories is related is more difficult. In
ref.\ \cite{Brandt:2020sho} this relationship is derived for the first
and the second order form of YM theory. Green's functions involving
the field strength in the first order formalism are related to Green's
functions involving composite fields in the second order
formalism. This was also done in the diagonal formulation of the first
order formalism of the YM \cite{Brandt:2018avq} and EH
\cite{Brandt:2016eaj} actions as shown in more detail in ref.\
\cite{Brandt:2020vre}. The renormalization of the first order
formalism of YM was studied in ref.\ \cite{Frenkel:2017xvm} and of the
diagonal formulation of first order 
YM in Ref.\ \cite{Brandt:2018lbe}. 
An alternative approach to a first order formulation of the YM theory
was given in Ref.~\cite{LAVROV2021136182r1}.

A second way of simplifying the computation of radiative effects in gauge theories is to use a Lagrange multiplier field in order to restrict the path integral used to quantize to paths that satisfy the classical Euler-Lagrange equations of motion. Unexpectedly, one finds in addition to the usual tree level effects, twice the usual one-loop contributions, and that all radiative effects beyond one-loop order vanish. The LM field participates in the gauge transformation and in associated Becchi--Rouet--Stora--Tyutin (BRST) transformation. This is considered in more detail in YM theory in ref.\ \cite{McKeon:1992rq} and with the EH action in \cite{Brandt:2018lbe, PhysRevD.100.125014}. The FP quantization procedure can be adapted to the presence of LM fields. In this paper we consider carefully how the LM field can be introduced and show how all radiative effects can be derived either by directly evaluating the Feynman path integral, and also by perturbative evaluation of Feynman diagrams. In the latter approach, the peculiar nature of the propagators for the physical fields and the LM fields plays a crucial role in eliminating all Feynman diagrams beyond one-loop order. Moreover we relate, in the LM framework, the Green functions in the first and second order formulations of the EH action.   

As a result of having no Feynman diagrams beyond one-loop order, it is possible to compute Green's function exactly. For YM theory, the exact relation between the bare and renormalized coupling can therefore be determined and the exact renormalization group functions can be found \cite{McKeon:1992rq}. With the EH action, it follows that all divergences (i.e.\ those arising at one-loop order) can be absorbed by a renormalization of the LM field \cite{Brandt:2018lbe, PhysRevD.100.125014}. Unitarity is retained; this follows from the BRST invariance of the full effective action and has been explicitly verified.

In the Appendix A, it is shown that after a LM field has been introduced, a redefinition of the fields eliminates mixed propagators and results in there being Feynman diagrams beyond one-loop order. However, these higher loop diagrams sum to zero since in the original formulation higher loop diagrams do not occur; this cancellation is verified at two-loop order in the Appendix B.

\section{First order Yang-Mills theory}
The YM Lagrangian in second order form 
\begin{equation}\label{eq:21}
    \mathcal{L}^{(2)}_{\text{YM}} = - \frac{1}{4} f^{a}_{\mu \nu} f^{a \, \mu
    \nu},
\end{equation}
where $f^{a}_{\mu \nu} \equiv  \partial_{\mu } {A}^{a}_{\nu} - \partial_{\nu}
{A}_{\mu }^{a} + g f^{abc} {A}_{\mu }^{b} {A}_{\nu }^{c}$, can be expressed in
first order form 
\begin{equation}\label{eq:22}
    \mathcal{L}^{(1)}_{\text{YM}} = \frac{1}{4} F^{a}_{\mu \nu} F^{a \, \mu
    \nu}
    - \frac{1}{2} F^{a}_{\mu \nu} f^{a \, \mu \nu}
\end{equation}
as can be seen classically by using the equation of motion 
\begin{equation}\label{eq:23}
    F^{a}_{\mu \nu} = {f}^{a}_{\mu \nu}.
\end{equation}
(An alternate first order formulation of the YM theory is based on the Duffin-Kemmer approach \cite{PhysRevD.20.462}.)
One can quantize the theory using $ \mathcal{L}^{(1)}_{\text{YM}} $
\cite{McKeon:1994ds, Brandt:2015nxa}; although one encounters two propagating fields $A^{a}_{\mu} $ and $
{F}^{a}_{\mu \nu} $, there is only one vertex $ -1/2 g f^{abc}
{F}^{b}_{\mu \nu}A^{b \, \mu} A^{c \, \nu } $ arising in
Eq.~\eqref{eq:22}.

Both
$  {F}_{\mu \nu}^{a}$ and $ {A}_{\mu}^{a} $ may appear on external legs of
Green's functions. The role of $ {F}_{\mu \nu}^{a} $ is discussed in detail in ref.\ \cite{Brandt:2020vre}. There the generating functional, when using the FP procedure,  
\begin{equation}\label{eq:24}
    Z[j,J] = \int \mathcal{D} {F}_{\mu \nu}^{a} \mathcal{D} A^{a}_{\mu} \exp
    \left [ i
    \int \mathop{d^{d}x} \left (  \frac{1}{4} F^{a}_{\mu \nu} F^{a \, \mu
    \nu}
- \frac{1}{2} F^{a}_{\mu \nu} f^{a \, \mu \nu} + \mathcal{L}_{\text{gf}}
+ j^{a}_{\mu} A^{a \, \mu} + {J}^{a}_{\mu \nu} {F}_{}^{a \, \mu \nu}
\right ) \right ] \Delta_{\text{FP}} (A)
\end{equation}
was subject to the shift
\begin{equation}\label{eq:25}
    {F}_{\mu \nu}^{a} = {\tilde{E}}_{\mu \nu}^{a} + {f}_{\mu \nu}^{a} - 2
    {J}_{\mu \nu}^{a}
\end{equation}
so that
\begin{eqnarray}\label{eq:26}
    Z[j,J] &=& \int \mathcal{D} {\tilde{E}}_{\mu \nu}^{a} \exp \left[i \int \mathop{d^{d}x} \left (\frac{1}{4} \tilde{E}^{a}_{\mu \nu} \tilde{E}^{a \, \mu\nu} \right) \right]
\nonumber \\ &&
    \int \mathcal{D} A^{a}_{\mu} \exp \left [i 
  \int \mathop{d^{d}x} \left (  
      {\mathcal{L}}_{\text{YM}}^{(2)}  + \mathcal{L}_{\text{gf}} + j^{a}_{\mu} A^{a \, \mu} + {J}^{a}_{\mu \nu} {f}_{}^{a \, \mu \nu} -
{J}_{\mu \nu}^{a} {J}_{}^{a \, \mu \nu}
\right ) \right ]\Delta_{\text{FP}} (A). 
\end{eqnarray}
The integral over $ {\tilde{E}}_{\mu \nu}^{a} $ in Eq.~\eqref{eq:26} results in an overall constant factor and can be discarded. If $ {J}_{\mu \nu}^{a} = 0$, so that there is no external field $ {F}_{}^{a \, \mu \nu} $, one is left with the generating functional for Green's function that uses $ {\mathcal{L}}_{\text{YM}}^{(2)} $. If $ {J}_{\mu \nu}^{a} \neq 0 $, in Eq.~\eqref{eq:24} it serves as a source for the auxiliary field $ {F}^{a \, \mu \nu}$ while in Eq.~\eqref{eq:26} it occurs as a source for the composite field $ {f}^{a \, \mu \nu}$ and also appears quadratically in order that both Eqs.~\eqref{eq:24} and \eqref{eq:26} can be used to compute the tree level contributions for the two-point functions $ \left \langle F F \right \rangle $.
In ref.\ \cite{Brandt:2020vre} it is shown that 
\begin{equation}\label{eq:27}
    \langle 0|T F^{a}_{\mu \nu} (x) A^{b_1}_{\alpha_{1}} (x_1) \cdots A^{b_n}
    _{\alpha_{n}}(x_n)| 0 \rangle 
= \langle 0|T f^{a}_{\mu \nu} (x) A^{b_1}_{\alpha_{1}} (x_1) \cdots A^{b_n}
_{\alpha_{n}} (x_n)| 0 \rangle
\end{equation}
follows from Eqs.~\eqref{eq:24} and \eqref{eq:26}. This is a realization of the classical equation Eq.~\eqref{eq:23} after quantization.

In Eq.~\eqref{eq:24}, there are tree level propagators $ \left \langle AA \right \rangle $, $ \left \langle AF \right \rangle $ and $ \left \langle FF \right \rangle $ and one three point vertex $(AAF)$, while in Eq.~\eqref{eq:26} there is one tree level propagator $ \left \langle AA \right \rangle $ and two vertices $(AAA)$ and $(AAAA)$. If one were to replace Eq.~\eqref{eq:25} with 
\begin{equation}\label{eq:28}
    {F}_{\mu \nu}^{a} = \tilde{H}_{ \mu \nu}^{a} + \partial_{\mu } {A}^{a}_{\nu} - \partial_{\nu} {A}_{\mu }^{a} - 2 {J}_{\mu \nu}^{a}
\end{equation}
then Eq.~\eqref{eq:24} becomes
\begin{equation}\label{eq:29}
    \begin{split}
        Z[j,J] &= \int \mathcal{D} {\tilde{H}}_{\mu \nu}^{a} \mathcal{D} A^{a}_{\mu} \exp i
        \int \mathop{d^{d}x} \bigg( \frac{1}{4} \tilde{H}^{a}_{\mu \nu} \tilde{H}^{a \, \mu
    \nu}
    - \frac{1}{4} (\partial_{\mu } {A}^{a}_{\nu} - \partial_{\nu}
    {A}_{\mu }^{a}) (\partial^{\mu } {A}^{a \, \nu} - \partial^{\nu}
    {A}^{a \, \mu }) \\ &
        - \frac{1}{2} g f^{abc} (\partial_{\mu } {A}^{a}_{\nu} - \partial_{\nu} {A}_{\mu }^{a}) {A}_{}^{b \, \mu} A^{c \, \nu } 
               - \frac{1}{2} g f^{abc} {\tilde{H}}_{\mu \nu}^{a} A^{b \, \mu} A^{c \, \nu} +  \mathcal{L}_{\text{gf}} + j^{a}_{\mu} A^{a \, \mu} +  {J}^{a}_{\mu \nu}  \tilde{H}^{a \, \mu \nu}  - {J}_{\mu \nu}^{a} {J}_{}^{a \, \mu \nu}
           \bigg) \Delta_{\text{FP}} (A).
    \end{split}
\end{equation}
There now are two propagating fields $ \tilde{H}^{a}_{\mu \nu} $ and $ A^{a}_{\mu} $, with no off diagonal propagators, but two vertices, $(AAA)$ and $(\tilde{H}AA)$. Again, Eq.~\eqref{eq:27} can be shown to follow from Eqs.~\eqref{eq:24} and \eqref{eq:29}.

\section{First Order Einstein-Hilbert action}
When one replaces the metric $ g_{ \mu \nu } $ and the affine connection $
{\Gamma}_{\mu \nu}^{\lambda} $ with \cite{Brandt:2015nxa} 
\begin{subequations}
    \begin{equation}\label{eq:318a}
        h^{\mu \nu} = \sqrt{- g} g^{\mu \nu}, \quad h_{\mu \nu} = \frac{g_{\mu \nu}}{\sqrt{-g}} 
    \end{equation}
    and
    \begin{equation}\label{eq:318b}
        {G}_{\mu \nu}^{\lambda} = {\Gamma}_{\mu \nu}^{\lambda} - \frac{1}{2}
        \left ( \delta^{\lambda}_{\mu} {\Gamma}_{\nu \sigma}^{\sigma} +
        \delta^{\lambda}_{\nu} {\Gamma}_{\mu \sigma}^{\sigma} \right ),
    \end{equation}
\end{subequations}
the EH action becomes \cite{Brandt:2020sho} 
\begin{equation}\label{eq:319}
    S_{\text{EH}} =  \frac{1}{\kappa^{2}} \int \dd
    \mathcal{L}^{(1)}_{\text{EH}}, 
\end{equation}
where $ \kappa^{2} = 16 \pi G_N$ and 
\begin{equation}\label{eq:320}
    \begin{split}
        \mathcal{L}^{(1)}_{\text{EH}} & = h^{\mu \nu} \left (
    {G}_{ \mu \nu, \lambda}^{ \lambda} + \frac{1}{d-1} {G}_{ \mu \lambda}^
{\lambda} {G}_{ \nu \sigma}^{ \sigma} - {G}_{ \mu \sigma}^{ \lambda}
{G}_{ \nu \lambda}^{ \sigma} \right ) \\ 
                                        & = - {G}_{\mu
                                        \nu}^{\lambda} h^{\mu \nu}_{,
                                    \lambda} + \frac{1}{2 }
                                    {G}_{\mu \nu}^{\lambda}
                                    {M}_{\lambda}^{\mu \nu}
                                    {}_{\sigma}^{\pi \tau} (h)
                                    {G}_{\pi \tau}^{\sigma},
\end{split}
\end{equation}
where
\begin{eqnarray}\label{eq:36BM2016}
    M^{\mu\nu}_{\lambda}{}^{\pi\tau}_{\sigma}(h)   & = &
\frac{1}{2}\left[\frac{1}{d-1}\left( \delta^\nu_\lambda\delta^\tau_\sigma h^{\mu\pi}+
                                                \delta^\mu_\lambda\delta^\tau_\sigma h^{\nu\pi}+
                                                \delta^\nu_\lambda\delta^\pi_\sigma h^{\mu\tau}+
                                                \delta^\mu_\lambda\delta^\pi_\sigma h^{\nu\tau}
\right) 
 \right.  \nonumber \\  && - \left. 
\left( 
                                                \delta^\tau_\lambda\delta^\nu_\sigma h^{\mu\pi}+
                                                \delta^\tau_\lambda\delta^\mu_\sigma h^{\nu\pi}+
                                                \delta^\pi_\lambda\delta^\nu_\sigma h^{\mu\tau}+
                                                \delta^\pi_\lambda\delta^\mu_\sigma h^{\nu\tau}
\right) \frac{}{} \!\!\right]. 
\end{eqnarray}

The equation of motion for $ {G}_{\mu \nu}^{\lambda} $ leads to
\begin{equation}\label{eq:321}
    G_{\mu \nu}^{\lambda} = (M^{-1}){}_{\mu \nu}^{\lambda} {}^{\sigma}_{\pi \tau} (h){h}^{\pi \tau}_{,\sigma}
\end{equation}
where
\begin{equation}\label{eq:323a}
    (M^{-1})_{\mu \nu}^{\lambda} {}_{ \pi \tau}^{ \rho} (h) = -
    \frac{1}{2(d-2)} h^{\lambda \rho} h_{\mu \nu} h_{\pi \tau} +
    \frac{1}{4} h^{\lambda \rho} \left ( h_{\pi \mu} h_{\tau \nu} + h_{\pi \nu} h_{\tau \mu}\right ) - \frac{1}{4} \left ( h_{\tau \mu} \delta_{\nu}^{\rho} \delta_{\pi}^{\lambda} + h_{\pi \mu} \delta_{\nu}^{\rho} \delta_{\tau}^{\lambda} +  h_{\tau \nu} \delta_{\mu}^{\rho} \delta_{\pi}^{\lambda} +  h_{\pi \nu} \delta_{\mu}^{\rho} \delta_{\tau}^{\lambda}\right ) 
\end{equation}
so that 
\begin{equation}\label{eq:323}
    (M^{-1})_{\alpha \beta }^{\rho} {}_{ \mu \nu}^{ \lambda }M^{\mu
    \nu}_{\lambda} {}^{ \pi \tau }_{ \sigma}
    = \Delta_{\alpha \beta}^{\pi \tau} \delta^{\rho}_{\sigma} \equiv
    \frac{1}{2} \left ( \delta_{\alpha}^{\pi} \delta_{\beta}^{\tau}
    +\delta_{\alpha}^{\tau} \delta_{\beta}^{\pi}  \right )
    \delta^{\rho}_{\sigma}.
\end{equation}
Eliminating $ {G}_{\mu \nu}^{\lambda} $ in $ {\mathcal{L}}_{\text{EH}}^{(1)}$ using Eq.~\eqref{eq:321} leads to the usual second order EH Lagrangian 
\begin{equation}\label{eq:322}
    \mathcal{L}^{(2)}_{\text{EH}} = - \frac{1}{2} h^{\mu \nu}_{, \lambda}
    ( M^{-1})^{\lambda}_{\mu \nu} {}_{\pi \tau}^{\sigma}
    h^{\pi \tau}_{, \sigma},
\end{equation}
Just as $ {\mathcal{L}}_{\text{YM}}^{(1)} $ and $ {\mathcal{L}}_{\text{YM}}^{(2)} $ are classically equivalent, $ \mathcal{L}^{(1)}_{\text{EH}} $ and $
\mathcal{L}^{(2)}_{\text{EH}} $ are classically equivalent. We now will establish the relationship between the generating functionals that follow from $ {\mathcal{L}}_{\text{EH}}^{(1)} $ and $
{\mathcal{L}}_{\text{EH}}^{(2)} $.

In using the path integral to quantize the EH action, it is necessary
to employ background field quantization \cite{DeWitt:1967uc, Abbott:1980hw, abbott82}
Taking the background metric to be flat ($ \eta^{\mu \nu} $),  
\begin{equation}\label{eq:324}
    {h}_{}^{\mu \nu} (x) = \eta^{\mu \nu} + \kappa \phi^{\mu \nu} (x)
\end{equation}
so that
\begin{equation}\label{eq:325}
    M(h) = m + \kappa M (\phi ),
\end{equation}
where $ m \equiv M(\eta)$, then the generating functional is
\begin{equation}\label{eq:326}
    \begin{split}
        Z[j,J] & =   \int \mathcal{D} {G}_{\mu \nu}^{\lambda} \mathcal{D}
    {\phi}_{}^{\mu \nu} \exp  \bigg\{ i\int \mathop{d^{d}x}
        \bigg[\frac{1}{\kappa^{2}} \left (- \kappa {G}_{\mu \nu}^{\lambda} \phi^{\mu \nu}_{, \lambda} +
\frac{1}{2} {G}_{\mu \nu}^{\lambda} [m + \kappa M(\phi)]
    {}_{\lambda}^{\mu \nu} {}_{\sigma}^{\pi \tau} {G}_{\pi \tau}^{\sigma} \right)\\ & +
        {\mathcal{L} }_{\text{gf}}^{} ( \phi )
               +  j_{\mu \nu} \phi^{\mu \nu} + J^{\mu \nu}_{\lambda}
           {G}_{\mu \nu}^{\lambda} \bigg] \bigg\} {\Delta}_{\text{FP}}^{} (\phi).
    \end{split} 
\end{equation}

The auxiliary field $ {G}_{\mu \nu}^{\lambda} $ enters $ {\mathcal{L}}_{\text{EH}}^{(1)} $ quadratically while the auxiliary field $ F_{\mu \nu}^{a} $ enters $ {\mathcal{L}}_{\text{YM}}^{(1)} $ linearly. Nevertheless, one can make use of the analogues of Eqs.~\eqref{eq:25} and \eqref{eq:28} to simplify 
$Z[j,J]$ in Eq.~\eqref{eq:326}. We first make the shift
\begin{equation}\label{eq:327}
    {G}_{\mu \nu}^{\lambda} = {E}_{\mu \nu}^{\lambda} +
    (M^{-1})^{\lambda}_{\mu \nu} {}_{\pi \tau}^{\sigma} (h)(
    {h}_{, \sigma}^{\pi \tau} - \kappa^2 
    {J}_{\sigma}^{\pi \tau} )
\end{equation}
so that Eq.~\eqref{eq:326} becomes
\begin{equation}\label{eq:328}
    \begin{split}
        Z[j,J] & =   \int \mathcal{D} {E}_{\mu \nu}^{\lambda} \mathcal{D}
    {\phi}_{}^{\mu \nu} \exp  \bigg\{ i \int \mathop{d^{d}x}
        \bigg[  \frac{1}{\kappa^{2}} \left (\frac{1}{2} {E}_{\mu \nu}^{\lambda} M_{\lambda}^{\mu \nu}
            {}_{\sigma}^{\pi \tau}(h) {E}_{\pi \tau}^{\sigma} - \frac{1}{2}
            (h^{\mu \nu}_{,\lambda}- \kappa^2 J^{\mu \nu}_{\lambda})
         (M^{-1})^{\lambda}_{\mu \nu}
         {}^{\sigma}_{\pi \tau}(h) (h^{\pi \tau}_{,\sigma}-\kappa^2J^{\pi
\tau}_{\sigma}) \right )\\ & +
{\mathcal{L} }_{\text{gf}}^{} 
               +  j_{\mu \nu} \phi^{\mu \nu} + J^{\mu \nu}_{\lambda}
       {G}_{\mu \nu}^{\lambda} \bigg] \bigg\} {\Delta}_{\text{FP}}^{} (\phi).
    \end{split} 
\end{equation}
Although Eqs.~\eqref{eq:25} and \eqref{eq:327} are analogous,
Eqs.~\eqref{eq:26} and \eqref{eq:328} differ in several ways. In Eq.~\eqref{eq:26} integration over $ \tilde{{E} }_{\mu \nu}^{a} $ leads to an inconsequential constant factor, while in
Eq.~\eqref{eq:328} integration over $ {E}_{\mu \nu}^{\lambda} $ results in a factor of
\begin{equation}\label{eq:329}
    \det{}^{-1/2} [m + \kappa M(\phi )].
\end{equation}
However, when this factor is expanded in powers of $ \kappa $, we end up with a sum of massless tadpole diagrams which go to zero when using dimensional regularization \cite{RevModPhys.47.849} and so the contribution of this determinant can be ignored.

Next, we consider the factor 
\begin{equation}\label{eq:330}
h^{\mu \nu}_{,\lambda}(M^{-1})^{\lambda}_{\mu \nu}{}^{\sigma}_{\pi
\tau}(h) h^{\pi \tau}_{,\sigma} = \kappa^{2}\phi^{\mu \nu}_{,
\lambda} (M^{-1})_{\mu \nu}^{\lambda} {}_{\pi \tau}^{\sigma} (\eta + \kappa
\phi) \phi^{\pi \tau}_{, \sigma}
\end{equation}
which is just proportional to $ {\mathcal{L}}_{\text{EH}}^{(2)} $; if we use the expansion 
\be\label{eq:331}
M^{-1}(\eta+\kappa\phi)=m^{-1}
-\kappa m^{-1} M(\phi) m^{-1}  
+\kappa^2 m^{-1} M(\phi) m^{-1} M(\phi)  m^{-1}  
+ \cdots,
\ee
then this term generates the infinite series of vertices that occurs in the second form of the EH action.

We also see that in $Z[j,J]$ of Eq.~\eqref{eq:328}, we have 
\begin{equation}
    \label{eq:25a}
\kappa {J}_{\lambda}^{\mu \nu} (M^{-1})_{\mu \nu}^{\lambda} {}_{\pi \tau}^{\sigma} \phi^{\pi \tau}_{, \sigma} 
\end{equation}
showing that $ {J}_{\lambda}^{\mu \nu} $ is the source for $ \kappa (M^{-1})_{\mu \nu}^{\lambda} {}_{\pi \tau}^{\sigma} \phi^{\pi \tau}_{, \sigma}$, which is consistent with Eq.~\eqref{eq:321}. It is analogous to the term $ J^{a}_{\mu \nu} f^{a \, \mu \nu} $ in Eq.~\eqref{eq:26}. 
Finally the term 
\begin{equation}
     J^{\mu \nu}_{\lambda}
{( M^{-1})}{}_{\mu \nu}^{\lambda} {}_{\pi \tau}^{\sigma}(h) {J}_{\sigma}^{\pi
\tau } 
\end{equation}
in Eq.~\eqref{eq:328} shows that the source $ J^{\mu \nu}_{\lambda} $ enters $Z[j,J]$ in a more complex way than $ {J}_{\mu \nu}^{a} $ enters into Eq.~\eqref{eq:26}. If $J=0$, we have shown that $Z[j,0]$ in Eq.~\eqref{eq:328} reduces to the generating functional for the second order EH action.

In place of the shift of Eq.~\eqref{eq:327}, let us consider 
\begin{equation}\label{eq:332}
    {G}_{\mu \nu}^{\lambda} = {H}_{\mu \nu}^{\lambda} + (m^{-1})
    {}_{\mu \nu}^{\lambda} {}_{\pi \tau}^{\sigma} \left ( \kappa \phi^{\pi \tau}
    _{, \sigma} - \kappa^2J^{\pi \tau}_{\sigma} \right ).
\end{equation}
Like Eq.~\eqref{eq:28}, this shift eliminates mixed propagators that occur in
Eq.~\eqref{eq:326}. We find that
\begin{equation}\label{eq:333}
    \begin{split}
Z[j,J] 
        & =   \int \mathcal{D} {H}_{\mu \nu}^{\lambda} \mathcal{D}
    {\phi}_{}^{\mu \nu} \exp  \biggm\{ i \int d^{d}x
    \biggm[   \frac{1}{\kappa^2} \biggm(\frac{1}{2} {H}_{\mu \nu}^{\lambda} [m + \kappa
            M(\phi)]_{\lambda}^{\mu \nu}
            \mbox{}_{\sigma}^{\pi \tau} {H}_{\pi \tau}^{\sigma} 
           \\
           &+ {H}_{\mu \nu}^{\lambda} [\kappa M\mbox{}_{\lambda}^{\mu \nu}
            \mbox{}_{\sigma}^{\pi \tau} (\phi )] (m^{-1}){}_{\pi \tau}^{\sigma}
           \mbox{}_{\alpha \beta}^{\rho} (\kappa {\phi}_{, \rho}^{\alpha \beta} -\kappa^2
            {J}_{\rho}^{\alpha \beta} )
            \\
&+ 
\frac{1}{2}
            (\kappa \phi^{\mu \nu}_{,\lambda}- \kappa^2 J^{\mu \nu}_{\lambda})
            [m^{-1} \kappa M(\phi ) m^{-1} - m^{-1}]\mbox{}^{\lambda}_{\mu \nu}
         \mbox{}^{\sigma}_{\pi \tau}(\kappa \phi^{\pi \tau}_{,\sigma}- \kappa^2 J^{\pi
    \tau}_{\sigma}) \biggm)\\
    &+
{\mathcal{L} }_{\text{gf}}^{}(\phi) 
               +  j_{\mu \nu} \phi^{\mu \nu} + J^{\mu \nu}_{\lambda}
           {G}_{\mu \nu}^{\lambda} \biggm] \biggm\}  {\Delta}_{\text{FP}}^{} (\phi).
    \end{split} 
\end{equation}

In Eq.~\eqref{eq:333} we find that the propagators for $ H $ and $ \phi $ are
decoupled, the former coming from the term $ 1/(2\kappa^2) H^{\lambda}_{\mu \nu} m
{}_{\lambda}^{\mu \nu} {}_{\sigma}^{\pi \tau} H^{\sigma}_{\pi \tau}$, the
latter form the terms $ - 1/2\, \phi^{\mu \nu}_{, \lambda} (m^{-1})
{}_{\mu \nu}^{\lambda} {}_{\pi \tau}^{\sigma} \phi^{\pi \tau}_{, \sigma} +
{\mathcal{L}}_{\text{gf}}^{} (\phi ) $. There are now three point vertices $(H
\phi H)$, $(H \phi \phi)$ and $(\phi \phi \phi )$ but one avoids the infinite series of vertices arising from Eq.~\eqref{eq:331} when it is inserted into Eq.~\eqref{eq:328}. 
The source $ {J}_{\lambda}^{\mu \nu} $ enters the action Eq.~\eqref{eq:333} both linearly and quadratically.

We see that the different forms of the generating functional $Z[j,J]$ given in Eqs.~\eqref{eq:326}, \eqref{eq:328} and Eq.~\eqref{eq:333} each have their respective advantages. These are analogous to the forms of $Z[j,J]$ given by Eqs.~\eqref{eq:24}, \eqref{eq:26} and \eqref{eq:29} for YM theory.

We now will consider the use of Lagrange multiplier fields to restrict the field configuration to solutions of the classical equations of motion when performing the path integral.

\section{Lagrange Multiplier Fields}

We first will illustrate how LM fields can be used in conjunction with the path integral. If we have a simple Lagrangian for a scalar field $ A(x) $ in the presence of a source $J(x)$ 
\begin{equation}\label{eq:429}
    {\mathcal{L}}_{\text{A}}^{} = \frac{1}{2} (\partial_{\mu} A)^2  - \frac{g}{3!} A^3 + J A
\end{equation}
then a LM field $B(x)$ can be used to ensure that the equations of motion are satisfied so that we have the action 
\begin{equation}\label{eq:430}
    {\mathcal{L}}_{\text{AB}}^{} = \frac{1}{2} (\partial_{\mu} A)^2  - \frac{g}{3!} A^3 + B \left ( - \partial^{2} A - \frac{g}{2} A^{2} + J \right ) + J A.
\end{equation}
If we consider the path integral 
\begin{equation}\label{eq:431}
    Z[J] = \int \mathcal{D}A \mathcal{D} B \exp i \int \mathop{d^{d}x} {\mathcal{L}}_{\text{AB}}^{}
\end{equation}
we see that with this form of the generating functional $Z$, Feynman diagrams cannot be generated as there is no source for the field $B$. However, one can integrate over B so that 
\begin{equation}\label{eq:432}
    Z[J] = \int \mathcal{D}A \mathop{\delta}( - \partial^{2} A - \frac{g}{2} A^{2} + J  ) \exp i \int \mathop{d^{d}x} \left ( \frac{1}{2} (\partial_{\mu} A)^2 - \frac{g}{3!} A^{3} + JA \right ).
\end{equation}

The $ \delta $-function now makes it possible to integrate over $A(x)$ by using the formula
\begin{equation}\label{eq:433}
    \int \mathop{\mathop{d^{d}x_i}} f(x_i) \mathop{\delta}\left( \frac{\partial g}{\partial x_i} (x_i)\right)= 
    \sum_{ \bar{x}_{i} }^{} \frac{f ( \bar{x}_{i} ) }{\left | \det \frac{\partial^{2} g }{\partial x_i  \partial x_j} ( \bar{x}_i)\right |},
\end{equation}
where $\bar{x}_{i}$ satisfies $ \frac{\partial }{\partial x_i}  g( \bar{x}_{i}) = 0$. Eq.~\eqref{eq:432} then becomes 
\begin{equation}\label{eq:434}
    Z[ J] = \sum_{\bar{A}_{i}}^{} {\left | \det(- \partial^{2} - g \bar{A} )\right |^{-1}}
\exp i  \int
\mathop{\mathop{d^{d}x}} \left ( \frac{1}{2} ( \partial_{\mu} \bar{A})^{2} - \frac{g}{3!} \bar{A}^{3} + J \bar{A}\right),
\end{equation}
where 
\begin{equation}\label{eq:435}
    - \partial^{2} \bar{A} - \frac{g}{2} \bar{A}^{2} + J  = 0
\end{equation}
defines $ \bar{A} [J] $. A perturbation solution to Eq.~\eqref{eq:435} for $ \bar{A} [J] $ in powers of $g$ is (with $ \partial^{2} D(x-y) \equiv \partial^{2} D_{xy} = \delta (x-y)$) 
\begin{equation}\label{eq:436}
        \bar{A} (x) =  D_{xy}J_y - \frac{g}{2} D_{xy} D_{yz_1}D_{y z_2} J_{z_1} J_{z_2} + \frac{g^{2}}{2} D_{xy} D_{yz_1} J_{z_1} D_{yz} D_{zz_2} J_{z_2} 
        D_{zz_3}J_{z_3} + \mathcal{O} (g^3),
\end{equation}
where we used the DeWitt notation for integration over repeated continuous indexes: $D_{xy} J_y \equiv \int d^d y D(x-y)J(y)$. 
Eq.~\eqref{eq:436} can be represented graphically by the tree graphs of Fig.\ 1.
\begin{figure}
    \includegraphics[scale=0.5]{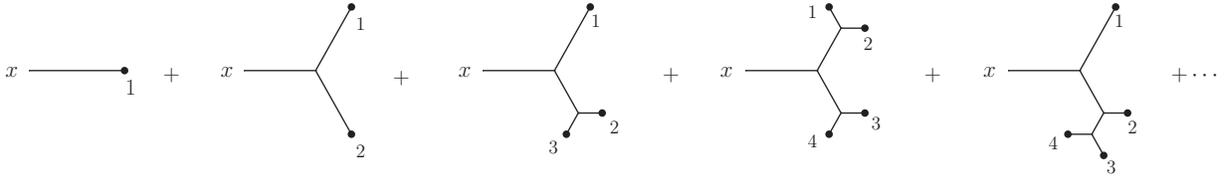}
    \label{fig:fig1}
    \caption{Graphical representation of $ \bar{A}[J]$.}
\end{figure}

The exponential in Eq.~\eqref{eq:434}, by ref.\ \cite{Boulware:1968zz}
represents the sum of all tree graphs to the generating functional if
one were to use the classical action $ {\mathcal{L}}_{\text{A}}^{} $
of Eq.~\eqref{eq:429}. The functional determinant is the square of the
functional determinant that arise if one were to start from the
classical action of Eq.~\eqref{eq:429} and have all classical fields
satisfying Eq.~\eqref{eq:435}. One could introduce a complex Bosonic scalar field $c$ and write 
\begin{equation}\label{eq:437}
    \det( - \partial^{2}  - g \bar{A} )^{-1} = \int \mathcal{D}c \mathcal{D} \bar{c} \exp i \int \mathop{d^{d}x} \bar{c} ( - \partial^{2}  - g \bar{A} ) c.
\end{equation}
We then see that Eq.~\eqref{eq:434} can be represented graphically by
\begin{equation}\label{eq:438}
    Z[J]=\vcenter{\hbox{\includegraphics[width=0.87\textwidth]{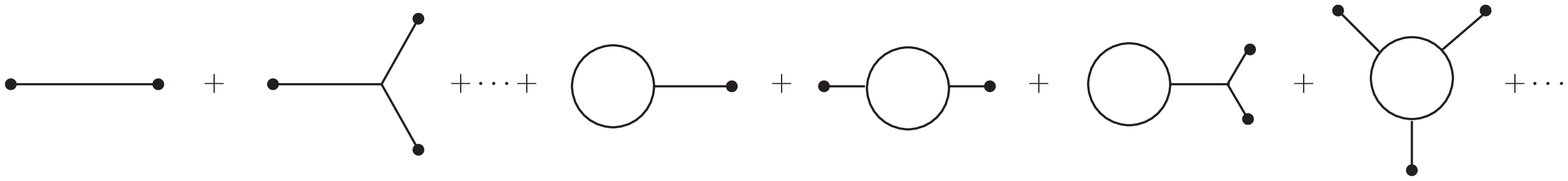}}}.
\end{equation}
In Eq.~\eqref{eq:438} we have only displayed connected graphs; disconnected graphs also contribute to $Z[J]$.
Green's function are found by computing 
\begin{equation}\label{eq:439}
    \langle 0|T A(x_1) \cdots A(x_n) | 0 \rangle = (-i)^{n} \frac{\delta^{n} }{\delta J(x_1) \cdots \delta J(x_n)} Z[J] \bigg|_{J=0}
\end{equation}
and we recover the tree level diagrams and twice the one-loop diagrams associated with $ {\mathcal{L}}_{\text{A}}^{} $ of Eq.~\eqref{eq:429} alone. No higher loop diagrams arises with $Z[J]$ in Eq.~\eqref{eq:434}; there are only the contributions of Eq.~\eqref{eq:438}.

A different approach to using the LM field is to consider in place of Eq.~\eqref{eq:430} 
\begin{equation}\label{eq:440}
    \tilde{{\mathcal{L}}}_{\text{AB}}^{} = \frac{1}{2} (\partial_{\mu} A)^2  - \frac{g}{3!} A^3 + B \left ( - \partial^{2} A - \frac{g}{2} A^{2} \right )
\end{equation}
and have sources for $A(x)$ and $B(x)$ so that 
\begin{equation}\label{eq:441}
    Z[J,K] = \int \mathcal{D}A \mathcal{D} B \exp i \int \mathop{d^{d}x} \left ( \tilde{{\mathcal{L}}}_{\text{AB}}^{} + JA + KB\right ) .
\end{equation}
Feynman rules can now be derived from Eq.~\eqref{eq:441} by writing it as
\begin{equation}\label{eq:442}
    \begin{split}
        Z[J,K] &= \sum_{n=0}^{ \infty } \frac{1}{n!} \left \{  \int \mathop{d^{d}x} \left [ - \frac{g}{3!} \left ( \frac{1}{i} \frac{\delta }{\delta J(x)} \right )^{3}\right ] - \frac{g}{2} \left ( \frac{1}{i} \frac{\delta }{\delta K(x)} \right ) \left ( \frac{1}{i} \frac{\delta }{\delta J(x)} \right )^{2}\right \}^{n} 
            \\ &\times \exp i \int \mathop{d^{d}x} \left [ 
            - \frac{1}{2} \begin{pmatrix}
                A & B
            \end{pmatrix}
            \begin{pmatrix}
                \partial^{2} & \partial^{2} \\
                \partial^{2} & 0
            \end{pmatrix}
            \begin{pmatrix}
                A \\ B
            \end{pmatrix}
            + \begin{pmatrix}
                J & K
            \end{pmatrix}
            \begin{pmatrix}
                A \\ B
            \end{pmatrix}
        \right ].
    \end{split}
\end{equation}
As
\begin{equation}\label{eq:443}
    \begin{pmatrix}
        \partial^{2} & \partial^{2} \\
        \partial^{2} & 0
    \end{pmatrix}^{-1}
    = 
    \begin{pmatrix}
        0 & \frac{1}{\partial^{2}}\\ 
        \frac{1}{\partial^{2}} & - \frac{1}{\partial^{2}}
    \end{pmatrix}
\end{equation}
we see from Eqs.~\eqref{eq:441} and \eqref{eq:442} that we have the Feynman rules of Fig.\ 2.
\begin{figure}
    \includegraphics[scale=0.8]{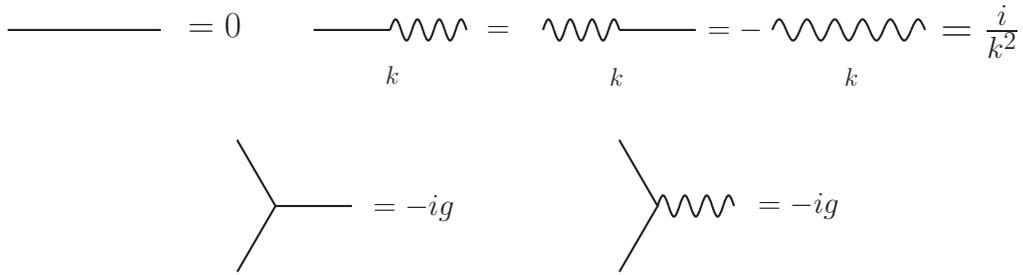}
    \label{fig:fig2}
    \caption{Feynman Rules from Eq.~\eqref{eq:441}. The quanta of the fields $A$ and $B$ are represented, respectively, by solid lines and wavy lines.}
\end{figure}
When drawing Feynman diagrams when external legs are included, it is possible to see with these Feynman rules that only tree and one-loop diagrams arise with tree diagrams having at most one external field $A$ and any number of external field $B$ while one-loop diagrams can only have external $B$ fields (when external propagators are not amputated). For example, we have the tree level diagram of Fig.\ 3 
\begin{figure}
    \includegraphics[scale=0.4]{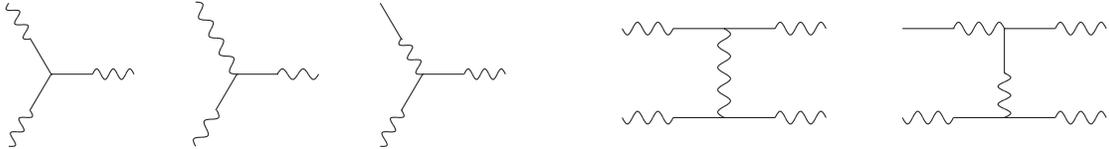}
    \caption{Tree diagrams.}
\end{figure}
and the one-loop diagrams of Fig.\ 4.
\begin{figure}
    \includegraphics[scale=0.5]{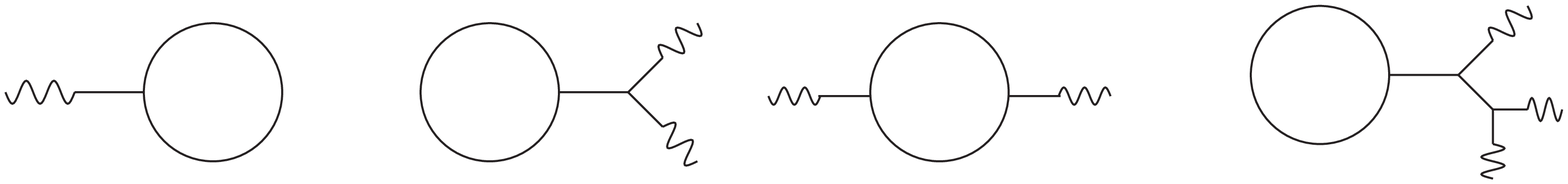}
    \caption{One-loop diagrams.}
\end{figure}

These diagrams can also be derived by performing the functional integration in Eq.~\eqref{eq:441}. First integrating over $B(x)$ we get 
\begin{equation}\label{eq:444}
    Z[J,K] = \int \mathcal{D}A \mathop{\delta } \left ( - \partial^{2} A - \frac{g}{2} A^{2} + K \right ) \exp i \int \mathop{d^{d}x} \left ( \frac{1}{2} ( \partial_{\mu} A )^{2} - \frac{g}{3!} A^{3} + JA \right )
\end{equation}
and then by Eq.~\eqref{eq:433}, we again arrive at Eq.~\eqref{eq:434} except now $ \bar{A} $ is replaced by $ \Bar{\Bar{{A}}} $ where in place of Eq.~\eqref{eq:435} we have 
\begin{equation}\label{eq:445}
    - \partial^{2} \bar{\bar{A}} - \frac{g}{2} \bar{\bar{A}}^{2} + K =0. 
\end{equation}
If now we want a Green's function with $m$ external fields $A$ and $n$ external fields $B$ we need to consider 
\begin{equation}\label{eq:446}
(-i)^{m} \frac{\delta^{m} }{\delta J(x_1) \cdots \delta J(x_m)} 
(-i)^{n} \frac{\delta^{n} }{\delta K(y_1) \cdots \delta K(y_n)} Z[J,K] 
\end{equation}
at $J=K=0$. 
However, in Eq.~\eqref{eq:444} the source $J$ only enters linearly in the exponential and by Eq.~\eqref{eq:445}, $ \bar{\bar{A}} $ depends only on $K$ and not on $J$, so we again see that one can have at most a single external $A$ field on a connected tree diagram with all other external fields being $B$ fields when using the generating functional $Z[J,K]$ in Eq.~\eqref{eq:441}. This is consistent with the Feynman diagrams of Figs. 3 and 4.

We should note though that the tree diagrams with a single external field $A$ on an unamputated external line (e.g.\ the third diagram of Fig.\ 3) are equal in magnitude but opposite in sign to certain graphs with only $B$ fields externally.  This is because of the difference in sign between the $ \left \langle AB \right \rangle $ and $ \left \langle BB \right \rangle $ propagator of Fig.\ 2. For example, the second and third diagrams of Fig.\ 3 have this property. Consequently, one finds that when all external legs of both tree level and one-loop diagrams in Figs. 3 and 4 are amputated, one is left with amplitudes that are identical to the amputated diagrams that from Eq.~\eqref{eq:438}.

The generating functional for connected diagrams is (restoring powers of $\hbar$)
\cite{Abbott:1980hw,abbott82}
\be\label{eqA}
W[J,K] = -i\hbar \ln Z[J,K].
\ee
If we define background fields $\bar A$, $\bar B$,
\begin{subequations}
\be\label{eqB}
\bar A = \frac{\delta W[J,K]}{\delta J}
\ee 
\be\label{eqC}
\bar B = \frac{\delta W[J,K]}{\delta K}
\ee 
\end{subequations}
then the Legendre transform
\be\label{eqD}
\Gamma[\bar A, \bar B] = W[J,K] - \int \dd (\bar A J + \bar B K)
\ee
(so that
\begin{subequations}
\be\label{eqE}
\bar J = -\frac{\delta \Gamma[\bar A,\bar B]}{\delta \bar A}
\;\;\;\;\;\;\;\; \left. \right.
\ee 
\be\label{eqF}
\bar K = -\frac{\delta \Gamma[\bar A,\bar B]}{\delta \bar B}
\;\;\;\;\;\;\;\; \left. \right)
\ee 
\end{subequations} 
generate the one-particle irreducible diagrams \cite{Abbott:1980hw,abbott82}
\be\label{eqG}
\langle 0| {\rm T} A(x_1)\cdots A(x_m) B(y_1)\cdots B(y_n)
\rangle_{{\rm 1PI}} 
= (\hbar)^{m+n} \frac{\delta^{m+n} \Gamma[\bar A, \bar B]}{\delta\bar
  A(x_1)\cdots \delta\bar A(x_m) \delta\bar B(y_1)\cdots \delta\bar B(y_n)}.
\ee

From Eqs. \eqref{eq:441}, \eqref{eqA}, \eqref{eqE} and \eqref{eqF} we
see that
\be\label{eqH}
{\rm e}^{{\frac i \hbar}\Gamma[\bar A, \bar B]} = \int {\cal D} A {\cal D} B
\exp{\frac i\hbar \int \dd\left(
\bar{\cal L}(\bar A+A,\bar B+B) 
- A \frac{\delta\Gamma[\bar A,\bar B]}{\delta\bar A}
- B \frac{\delta\Gamma[\bar A,\bar B]}{\delta\bar B}
\right)}.
\ee
The expansion
\be\label{eqI}
\Gamma[\bar A, \bar B] = \Gamma_0[\bar A, \bar B]+ \hbar \Gamma_1[\bar A, \bar B]+ \hbar^2\Gamma_2[\bar A, \bar B]+\cdots 
\ee
and
\be\label{eqJ}
\bar {\cal L}(\bar A+A, \bar B+B) = \left[
1+\frac{1}{1!}\left(A\frac{\partial}{\partial\bar A} +B\frac{\partial}{\partial\bar B} \right)+
\frac{1}{2!}\left(A^2\frac{\partial^2}{\partial\bar A^2}+B^2\frac{\partial}{\partial\bar B^2} 
+2 A B\frac{\partial^2}{\partial\bar A\partial\bar
  B}\right)+\cdots\right]{\cal L}(\bar A,\bar B)
\ee
show that
\be\label{eqK}
\Gamma_0[\bar A,\bar B]=\int \dd \,\bar{\cal L}(\bar A,\bar B) 
\ee
and
\be\label{eqL}
\Gamma_1[\bar A,\bar B]=0
\ee
with terms linear in $A$ and $B$ consequently canceling. There is no
need for the background fields $\bar A$ and $\bar B$ to satisfy the
equations of motion.

If in place of the generating functionals of Eqs.~\eqref{eq:431} and \eqref{eq:441} we were to use 
\begin{equation}\label{eq:447}
    Z'[J,K] = \int \mathcal{D}A \mathcal{D} B \exp i \int \mathop{d^{d}x} \left ( \frac{1}{2} ( \partial_{\mu} A)^{2} - \frac{g}{3!} A^{3} + B (- \partial^{2} A - \frac{g}{2} A^2 + J) + JA + KB \right ) 
\end{equation}
then by setting 
\begin{equation}\label{eq:448}
    C = A + B
\end{equation}
we would have 
\begin{equation}\label{eq:449}
    Z'[J,K] = \int \mathcal{D}C \mathcal{D} B \exp i \int \mathop{d^{d}x} \left [ \frac{1}{2} [( \partial_{\mu} C)^{2} - ( \partial_{\mu} B)^{2}] - g\left(\frac{1}{3!} C^3 - \frac{1}{2!} CB^2 + \frac{2}{3!} B^3\right) + JC+ KB \right ]. 
\end{equation}
Integration over $B$ in Eq.~\eqref{eq:447} leads to a $ \delta $-function that shows that again there are no diagrams beyond one-loop, as can be confirmed by using the Feynman diagrams that follow from Eq.~\eqref{eq:447}. In Eq.~\eqref{eq:449} the propagators no longer have non-diagonal form; we have the Feynman rules of Fig.\ 5.
\begin{figure}
    \includegraphics[scale=0.8]{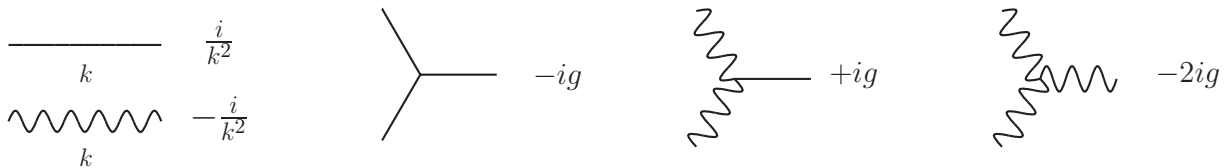}
    \caption{Feynman rules from Eq.~\eqref{eq:449}. The quanta of the field C is represented as a solid line.}
\end{figure}
With the vertices that follow for Eq.~\eqref{eq:449} it is evident that there are higher loop diagrams such as these of Fig.\ 6.
\begin{figure}[h]
    \includegraphics[scale=0.5]{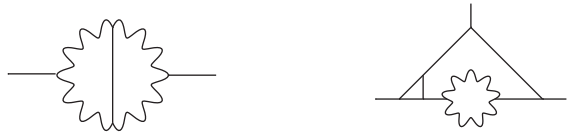}
    \caption{Higher loop diagrams from Eq.~\eqref{eq:449}.}
\end{figure}
The equivalence of Eqs.~\eqref{eq:447} and \eqref{eq:449} guarantees that such higher loop contributions sum to zero. (For an example, see Fig.\ 11 in Appendix B.) Functional derivatives of Eqs.~\eqref{eq:447} and \eqref{eq:449} with respect to $J$ and $K$ lead to the identities 
\begin{equation}\label{eq:450}
    \left \langle 0| T[A(x_1) + B(x_1)] \cdots [A(x_m) + B(x_m)]B(y_1) \cdots B(y_n) |0\right \rangle = \left \langle 0|T C(x_1) \cdots C ( x_{m} ) B(y_1) \cdots B(y_n) |0 \right \rangle.
\end{equation}
In the Appendix A is shown a generalization of Eq.~\eqref{eq:449}. Some examples of the identity Eq.~\eqref{eq:450} have been checked to one and two-loop order in Appendix B. 

It is important to remark that while Eq.~\eqref{eq:447} is linear in the LM field $ B$ which allows it to be immediately integrated which yields a $ \delta$-function. However, in Eq.~\eqref{eq:449} terms that are non-linear in $B$ appear and we cannot integrate the LM field in the same way. Nevertheless, Eq.~\eqref{eq:450} guarantees that all contributions beyond one-loop that follow from Eq.~\eqref{eq:449} vanish as well. Note that diagrammatically, using the Feynman rules of Fig.\ 5, one sees that at tree level the Green's functions with only external $C$ fields are the same of the Green's functions with only external $A$ fields that follow from the generating function that uses Eq.~\eqref{eq:429}, and provided we double the one-loop results. In addition Green's functions with only external $A$ fields resulting from Eq.~\eqref{eq:430} vanish at any order. 

It is also of interest to consider the Dyson-Schwinger \cite{Dyson:1949ha, Schwinger:1951ex} equation for the model defined by the generating functional of Eq.~\eqref{eq:441}. By making the shifts of integration variable 
\begin{equation}\label{eq:451}
    A \to A + \epsilon f_A(A,B), \quad B \to B + \epsilon f_B(A,B)
\end{equation}
in Eq.~\eqref{eq:441} and having lowest order terms set equal to zero we find that 
\begin{subequations}
    \label{eq:452}
\begin{equation}\label{eq:452a}
    \left \{ - \partial^{2} \left [ \left ( \frac{1}{i} \frac{\delta }{\delta J(x)}  \right ) + \left ( \frac{1}{i} \frac{\delta }{\delta K(x)}  \right )  \right ] - g \left [ \frac{1}{2} \left ( \frac{1}{i} \frac{\delta }{\delta J(x)}  \right )^{2} +  \left ( \frac{1}{i} \frac{\delta }{\delta K(x)}  \right )\left ( \frac{1}{i} \frac{\delta }{\delta J(x)}  \right )\right ] + J(x) \right \} Z[J,K]=0
\end{equation}
and 
\begin{equation}\label{eq:45b}
    \left [ - \partial^{2}  \left ( \frac{1}{i} \frac{\delta }{\delta J(x)}  \right ) -  \frac{g}{2} \left ( \frac{1}{i} \frac{\delta }{\delta J(x)}  \right )^{2} + K(x) \right ] Z[J,K]=0.
\end{equation}
\end{subequations}
These are the Dyson-Schwinger equations that are satisfied by $Z$. A perturbative expansion of $Z$ in powers of $g$ in Eq.~\eqref{eq:452} yields the results previously found by the Feynman diagrams that follow from Eq.~\eqref{eq:442}, or by the functional integration of Eqs.~\eqref{eq:435} and \eqref{eq:445}. An approach introduced by Mandelstam that is closely related to that of Schwinger and Dyson is given in ref.\ \cite{Mandelstam:1968hz}.

\section{Gauge Theories with Lagrangian Multiplier Fields}

When LM fields are used to ensure that the classical equations of motion are satisfied in a gauge theory, special care must be taken as the LM fields are themselves gauge fields \cite{PhysRevD.100.125014}. We will review some features of such gauge symmetries before examining how LM fields can be used in conjunction with the first order YM and EH actions.

If the action is invariant under 
\begin{equation}\label{eq:453}
    \phi'_{i} = \phi_{i} + H_{ij} (\phi ) \xi_{j}
\end{equation}
we find that 
\begin{equation}\label{eq:454}
     \int \mathop{\mathop{d^{d}x}}  H_{ij} (\phi )
     \frac{\partial {\mathcal{L} [\phi_{i} ]} }{\partial \phi_{i}} \xi_{j} =0.
\end{equation}
When the action is supplemented by a LM field, then
\begin{equation}\label{eq:455}
    S_{\text{LM}} = \int \mathop{\mathop{d^{d}x}} \left ( \mathcal{L}(\phi_i) + \lambda_{i} \frac{\partial
            {\mathcal{L}[\phi_i]}
    }{\partial \phi_{i}}\right )
\end{equation}
has the invariance of Eq.~\eqref{eq:453} supplemented by
\begin{equation}\label{eq:456}
    \lambda'_{i} = \lambda_{i} + \lambda_{l} \frac{\partial {H_{ij} (\phi )} }{\partial
    \phi_{l}} \xi_{j}
\end{equation}
as well as the invariance 
\begin{equation}\label{eq:457}
    \lambda'_{i} = \lambda_{i} + H_{ij} ( \phi ) \zeta_{j}.
\end{equation}

When using the standard FP procedure, if the action
\begin{equation}\label{eq:458}
    S = \int \mathop{\mathop{d^{d}x}} \mathcal{L} (\phi )
\end{equation}
has the invariance of Eq.~\eqref{eq:453} then the generating functional
\begin{equation}\label{eq:459}
    Z[j_i] = \int \mathcal{D} \phi_{i} \exp i \int \mathop{\mathop{d^{d}x}}
    \left ( \mathcal{L} (\phi ) + j_{i} \phi_{i} \right )
\end{equation}
is supplemented by a constant 
\begin{equation}\label{eq:460}
    \int \mathcal{D} \xi_{i} \delta \boldsymbol{(} F_{ij} ( \phi_{j} +H_{jk} (\phi )  \xi
    _{k}) - p_{i}\boldsymbol{)} \det F_{ij} H_{jk} (\phi )
\end{equation}
followed by 
\begin{equation}\label{eq:461}
    \int \mathcal{D} p_{i} \exp i \int \mathop{\mathop{d^{d}x}} \left (
    - \frac{1}{2 \alpha} p_{i} p_{i} \right )
\end{equation}
leaving us with
\begin{equation}\label{eq:462}
    Z[j_{i}] = \int \mathcal{D} \xi_{i} \int \mathcal{D} \phi_{i}  \exp i \int
    \mathop{\mathop{d^{d}x}} \left ( \mathcal{L} (\phi_i ) - \frac{1}{2 \alpha} (F_{ij} \phi_{j} )^{2}  + j_{i}\phi_{i}\right ) \det (F_{ij} H_{jk} (\phi))
\end{equation}
once the gauge transformation of Eq.~\eqref{eq:453} has been performed with $ \xi_{i} $ replaced by $ - \xi_{i} $. Under this transformation, $ \mathcal{D} \phi_{i} \det( F_{ij} H_{jk} (\phi ))$ and $ {\mathcal{L}} (\phi_{i} )$ are invariant. The factor $ \int \mathcal{D} \xi_{i} $ in Eq.~\eqref{eq:462} is a multiplicative constant. The functional determinant can be exponentiated using Fermionic ``ghost'' fields $c_i$ and $ \bar{c}_{i} $, 
\begin{equation}\label{eq:463}
    \det( F_{ij} H_{jk} (\phi )) = \int
    \mathcal{D} \bar{c}_{i}\mathcal{D} c_{i} \exp i \int
    \mathop{\mathop{d^{d}x}}  \bar{c}_{i} F_{ij} H_{jk} (\phi ) c_{k}.
\end{equation}
The requirement that
\begin{equation}\label{eq:464}
    F_{ij} \phi_{j} = 0
\end{equation}
is a ``gauge condition''. It is also possible to impose multiple gauge conditions on a gauge field \cite{Brandt:2007td}. This is particularly useful in a spin-two gauge theory if one desires to have a propagator that is both traceless and transverse. 

If we now were to consider the action $S_{\text{LM}} $ of Eq.~\eqref{eq:455} with the gauge invariances of Eqs.~\eqref{eq:454}, \eqref{eq:456} and \eqref{eq:457}, then if $ \phi_{i} $ satisfies the gauge condition of Eq.~\eqref{eq:464} and $ \lambda_{i} $ satisfies 
\begin{equation}\label{eq:465} 
    F_{ij} \lambda_{j} = 0
\end{equation}
then the FP procedure of Eqs.~\eqref{eq:460} and \eqref{eq:461} is adapted to lead to insertion of 
\begin{equation}\label{eq:466}
    \int \mathcal{D} \xi_{i} \mathcal{D} \zeta_{i} \delta \left \{ F_{ij}
        \left[
        \begin{pmatrix}
            \phi_{j} \\
            \lambda_{j} 
        \end{pmatrix}
        + 
        \begin{pmatrix}
            0 & H_{jk} \\ 
            H_{jk} & \lambda_{l} \frac{\delta {H_{jk}} }{\delta \phi_{l}
        }
        \end{pmatrix}
        \begin{pmatrix}
            \zeta_{k}     \\ 
            \xi_{k}
        \end{pmatrix}
    \right]
        - 
        \begin{pmatrix}
            p_i \\
            q_i
        \end{pmatrix}
        \right
    \} \det        \begin{pmatrix}
            0 & F_{ij} H_{jk} \\
            F_{ij} H_{jk} & F_{ij} \left (\lambda_{l} \frac{\delta
            {H_{jk}} }{\delta \phi_{l}} \right )
        \end{pmatrix}
\end{equation}
followed by 
\begin{equation}\label{eq:467}
    \int \mathcal{D} p_{i} \exp i \int \mathop{\mathop{d^{d}x}} \left (
    - \frac{1}{2 \alpha} (p_{i} p_{i} +2 p_{i} q_{i}) \right ).
\end{equation}
Upon replacing the determinant in Eq.~\eqref{eq:466} with
\begin{equation}
    \label{eq:468 }
    \det\begin{pmatrix}
            0 & F_{ij} H_{jk} \\
            F_{ij} H_{jk} & F_{ij} \left ( H_{jk} + \lambda_{l} \frac{\delta
            {H_{jk}} }{\delta \phi_{l}} \right )
        \end{pmatrix}
    \end{equation}
and exponentiating this determinant using Fermionic scalars $c_{i} $, $ {\bar{c}}_{i}^{} $, $ d_i$, $ \bar{d}_i$, the generating functional of the form of Eq.~\eqref{eq:441} becomes
\begin{equation}\label{eq:469}
    \begin{split}
        Z[j_{i}, k_i] & = \int \mathcal{D} \phi_{i} \mathcal{D} \lambda_{i}   
    \int \mathcal{D} \bar{c}_{i}\mathcal{D} c_{i} \int \mathcal{D} \bar{d}_{i} \mathcal{D} d_{i} 
    \exp i \int
    \mathop{\mathop{d^{d}x}} \bigg[ \mathcal{L} (\phi_i ) + \lambda_{i} \frac{\partial
        {\mathcal{L}} }{\partial \phi_{i}}  - \frac{1}{2 \alpha} \left[ (F_{im}
        \phi_{m} )(F_{i n} \phi_{n}) + 2 (F_{im} \phi_{m})(F_{in}
        \lambda_{n} )\right] \\ &  + {\bar{c} }_{i} F_{ij} \left( H_{jk} +
    \lambda_{l} \frac{\partial {H_{jk}} }{\partial \phi_{l}} \right) c_{k} +
    \bar{d}_{i} F_{ij} H_{jk} c_{k} + 
    \bar{c}_{i} F_{ij} H_{jk} d_{k} + j_{i}\phi_{i} + k_i \lambda_i \bigg].
    \end{split}
\end{equation}
It is apparent that in Eq.~\eqref{eq:469}, just as $ \lambda_{i} $ is a LM field associated with $ \phi_i $, $d_i$ and $ \bar{d}_i $ are LM multiplier fields associated with $c_i$, $ \bar{c}_i$.

For consistency, the commutator of two gauge transformations of the
form of Eq. \eqref{eq:453} must itself be a gauge transformation, and so 
\be\label{eq2A}
\left(\delta_1\delta_2-\delta_2\delta_1\right) \phi_i=\left(
\frac{\partial H_{i m}}{\partial\phi_l} H_{l n}-
\frac{\partial H_{i n}}{\partial\phi_l} H_{l m}
\right) \left(\xi_{2 m} \xi_{1n}\right)
\equiv \left(f_{m n, p} H_{ip}\right)\left(\xi_{2 m} \xi_{1n}\right).
\ee
The structure constants $f_{m n, p}$ must satisfy
\be\label{eq2C}
f_{q a,p} f_{b c,q}+f_{q c,p} f_{a b,q}+f_{q b,p} f_{c a,q} = 0
\ee
in order that
\be\label{eq2D}
\left[\delta_1,\left[\delta_2,\delta_3\right]\right]+
\left[\delta_2,\left[\delta_3,\delta_1\right]\right]+
\left[\delta_3,\left[\delta_1,\delta_2\right]\right] = 0.
\ee
By using Eq. \eqref{eq2C}, it is possible to show that
\begin{eqnarray}\label{eq2E}
\int d^d x\left[{\cal L}(\phi_i)+ \lambda_i\frac{\partial{\cal L}}{\partial\phi_i}
-\frac{1}{2\alpha} \left[
(F_{i m} \phi_m) (F_{i n} \phi_n)  + 2  (F_{i m} \phi_m) (F_{i n} \lambda_n)
\right] 
\right.\nonumber \\ \left.
+{\bar c}_i F_{ij}\left(H_{jk} + \lambda_l \frac{\partial H_{jk}}{\partial\phi_l}\right) 
+{\bar d}_i F_{ij}\ H_{jk}  c_k +{\bar c}_i F_{ij}\ H_{jk}  d_k \right] 
\end{eqnarray}
is invariant under the BRST transformations
\begin{subequations}
\be\label{eq2F}
\delta\phi_i = H_{ij} c_j \epsilon 
\ee
\be\label{eq2G}
\delta\lambda_i = \left(H_{ij} d_j  + \lambda_l\frac{\partial H_{ij}}{\partial\phi_l} c_j\right)\epsilon 
\ee
\be\label{eq2J}
\delta\bar{c_i} = -\frac{1}{\alpha} F_{ij} \phi_j \epsilon 
\ee
\be\label{eq2K}
\delta\bar{d_i} = -\frac{1}{\alpha} F_{ij} \lambda_j \epsilon 
\ee
\be\label{eq2L}
\delta{c_i} = \frac{1}{2}  f_{p q, i} c_p c_q\epsilon 
\ee
\be\label{eq2M}
\delta{d_i} = f_{p q, i} c_p d_q\epsilon 
\ee
\end{subequations}
 where $\epsilon$ is a Grassmann constant. If $Q$ is defined so that
 $\delta\phi_i = (Q\phi_i)\epsilon$ etc, then $Q^2=0$ provided
 $f_{ab,c}$ is independent of $\phi_i$.

\section{First Order Yang-Mills Theory with Lagrange Multipliers}

We now supplement the classical Lagrangian ${\cal L}_{\text{YM}}^{(1)}$ for
first order YM theory in Eq. \eqref{eq:22} with LM fields to ensure
that the classical equations of motion are satisfied,
\be\label{eq3A1}
{\cal L}_{\lambda \text{YM}}^{(1)} = \frac 1 4 F^a_{\mu\nu} F^{a\,\mu\nu}
-\frac 1 2 F^a_{\mu\nu} f^{a\,\mu\nu}
+\lambda^a_\nu D^{ab}_\mu F^{b\,\mu\nu}
+\Lambda^a_{\mu\nu}  \left(F^{a\,\mu\nu} - f^{a\,\mu\nu}\right).
\ee
This classical action can be quantized using the procedure outlined in
the preceding section.

The gauge invariance of Eq. \eqref{eq:453} for the Lagrangian of
Eq. \eqref{eq:22} is 
\be\label{eq3A}
\delta_\xi   A^a_\mu = D^{ab}_\mu(A) \xi^b 
\ee 
\be\label{eq3B}
\delta_\xi   F^a_{\mu\nu} = g f^{abc} F^b_{\mu\nu} \xi^c 
\ee 
where $D^{ab}_\mu = \partial_\mu\delta^{ab} + g f^{apb} A^p_\mu$.
Under these variations
\be\label{eq3C}
\int  \dd\left[\frac{\partial {\cal L}^{(1)}_{\text{YM}}}{\partial A^a_\mu}\delta_\xi A^a_\mu\right]=0=
\int  \dd\left[\frac{\partial {\cal L}^{(1)}_{\text{YM}}}{\partial F^a_{\mu\nu}}\delta_\xi F^a_{\mu\nu}\right]
\ee
which shows that under the variations 
\be\label{eq3D}
\delta_\zeta   \lambda^a_\mu = D^{ab}_\mu(A) \zeta^b 
\ee 
\be\label{eq3E}
\delta_\zeta   \Lambda^a_{\mu\nu} = g f^{abc} F^b_{\mu\nu} \zeta^c 
\ee 
the action $\int \dd {\cal L}^{(1)}_{\lambda \text{YM}}$ is left
invariant. This is the transformation of Eq. \eqref{eq:457}. In
addition, since if $A^{\prime\, a}_\mu=A^a_\mu+\delta_\xi A^a_\mu$, 
$F^{\prime\, a}_{\mu\nu}=F^a_{\mu\nu}+\delta_\xi F^a_{\mu\nu}$, 
then Eq. \eqref{eq3C} shows that 
\be\label{eq3F}
\int \dd\left(
    \lambda^a_\mu\frac{\partial{\cal L}^{(1)}_{\text{YM}}}{\partial A^a_\mu}+
\Lambda^a_{\mu\nu}\frac{\partial{\cal L}^{(1)}_{\text{YM}}}{\partial F^a_{\mu\nu}}\right)
=
\int \dd\left(
    \lambda^{\prime\, a}_\mu\frac{\partial{\cal L}^{(1)}_{\text{YM}}}{\partial A^{\prime\, a}_\mu}+
\Lambda^{\prime\, a}_{\mu\nu}\frac{\partial{\cal L}^{(1)}_{\text{YM}}}{\partial F^{\prime\, a}_{\mu\nu}}\right)
\ee
provided Eqs. \eqref{eq3A} and \eqref{eq3B} are accompanied by
\be\label{eq3G}
\lambda^{\prime\, a}_\mu-\lambda^{ a}_\mu=\delta_\xi \lambda^a_\mu = g f^{abc}\lambda^b_\mu\xi^c
\ee
\be\label{eq3H}
\Lambda^{\prime\, a}_{\mu\nu}-\Lambda^{ a}_{\mu\nu}=\delta_\xi
\Lambda^a_{\mu\nu} = g f^{abc}\Lambda^b_{\mu\nu}\xi^c .
\ee
Eqs. \eqref{eq3G} and \eqref{eq3H} are the transformations of Eq. \eqref{eq:456}.

It is also possible to derive the transformations of
Eqs. \eqref{eq3A}, \eqref{eq3B}, \eqref{eq3D}, \eqref{eq3E},
\eqref{eq3G} and \eqref{eq3H} by using the formalism of Ref. \cite{Henneaux:1990au}.

We consequently find that the action
\be\label{eq3I}
S^{(1)}_{\lambda \text{YM}} = \int \dd {\cal L}^{(1)}_{\lambda \text{YM}}
\ee
is invariant under gauge transformations of the fields $A^a_\mu$ and $\lambda^a_\mu$
of Eqs. \eqref{eq3A}, \eqref{eq3D} and \eqref{eq3G}.
If we were to quantize this model using the Faddeev--Popov procedure
\cite{Faddeev:1967fc} it is necessary to use a gauge fixing condition. The
general gauge choices of Eqs. \eqref{eq:464} and \eqref{eq:465} are
taken in this case to be
\be\label{eq3J}
\partial_\mu\lambda^{a\mu}=0=\partial_\mu A^{a\mu} ,
\ee
Eq. \eqref{eq:466} now becomes the constant
\be\label{eq3K}
\int {\cal D}\xi^a {\cal D}\zeta^a
\delta\left(
\begin{array}{l}
\partial_\mu\left(A^{a\mu}+ D^{ab\mu}(A) \xi^b\right)+p^b \\
\partial_\mu\left(\lambda^{a\mu}+ D^{ab\mu}(A) \zeta^b+g f^{abc}
  \lambda^{b\mu}\right)+ q^b
\end{array}\right)
\det\left(
\begin{array}{cc}
0 & \partial\cdot D^{ab}(A) \\ \partial\cdot D^{ab}(A) & g 
                                           f^{apb} \partial\cdot\lambda^p \xi^b 
\end{array}
\right).
\ee
When this is used in conjunction with the constant
\be\label{eq3L}
\int  {\cal D} p^a {\cal D} q^a {\rm e}^{-\frac{i}{2\alpha}\int \dd \left(p^a p^a+ 2 p^a q^a\right)}
\ee
then the path integral for the generating functional for our model
becomes
\begin{eqnarray}\label{eq3M}
    Z^{(1)}_{\lambda \text{YM}} = \int {\cal D} \xi^a {\cal D} \zeta^a \int {\cal D} A^a_\mu {\cal D} F^a_{\mu\nu}
\exp i\bigg(
    S^{(1)}_{\text{YM}} -\frac{1}{2\alpha} \int \dd \big[
(\partial \cdot A^{a})^2  + 2 (\partial \cdot A^{a})  (\partial \cdot \lambda^{a})
 \nonumber\\
+j^a_\mu A^{a\mu} +J^a_{\mu\nu} F^{a\mu\nu} + k^{a}_\mu\lambda^{a\mu} 
+ K^{a}_\mu\Lambda^{a\mu} 
\big]\bigg) \nonumber \\
\det\left(
\begin{array}{cc}
0 & \partial\cdot D^{ab}(A) \\ \partial\cdot D^{ab}(A) & g 
                                           f^{apb} \partial\cdot\lambda^p \xi^b 
\end{array}
\right).
\end{eqnarray}
Using the identity
\be\label{eq3N}
\det\left(\begin{array}{cc} 0 & a \\ a & b\end{array}\right) = \det\left(\begin{array}{cc} 0 & a \\ a & a+b\end{array}\right)
\ee
and exponentiating the determinant by use of the Fermionic ghost
fields $c^a$, ${\bar c}^a$, $d^a$, ${\bar d}^a$, Eq. \eqref{eq3M}
becomes Eq.  \eqref{eq:670} with the integral over the gauge functions
$\xi^a$, $\zeta^a$ being an overall multiplicative factor.
\begin{equation}\label{eq:670}
    \begin{split}
        Z[j,J,k,K] = & \int \mathcal{D} {F}_{\mu \nu}^{a} \mathcal{D} {A}_{\mu}^{a} \int \mathcal{D} {\Lambda}_{\mu \nu}^{a} \mathcal{D} {\lambda}_{ \mu}^{a} 
    \int \mathcal{D} \bar{c}^{a} \mathcal{D} c^{a} \int \mathcal{D }
    \bar{d}^{a} \mathcal{D} d^{a} \exp i \int \mathop{d^{d} x} \bigg( \frac{1}{4} F^{a}_{\mu \nu} F^{a \, \mu
    \nu}
    - \frac{1}{2} F^{a}_{\mu \nu} f^{a \, \mu \nu} + \lambda^{a}_{\nu} D^{ab}_{ \mu} F^{b \, \mu \nu} \\
                 & + \Lambda^{a}_{\mu \nu} (F^{a \, \mu \nu} - f^{a \, \mu \nu} )- \frac{1}{2 \alpha} (\partial^{\mu} A^{a}_{\mu} )^2 - \frac{1}{\alpha} \partial^{\mu} A^{a}_{\mu} \partial^{\nu} \lambda^{a}_{\nu} + \bar{c}^{a} \partial^{\mu} D^{ab}_{\mu} (A + \lambda ) c^{b} + \bar{c}^{a} \partial^{\mu} D^{ab}_{\mu} (A ) d^{b} + \bar{d}^{a} \partial^{\mu} D^{ab}_{\mu} (A) c^{b} \\
                 & + {j}_{\mu}^{a} A^{a \, \mu} + J^{a}_{\mu \nu} {F}_{}^{a \,
             \mu \nu} + k_{\mu}^{a} \lambda^{a \, \mu} + K_{\mu \nu}^{a} \Lambda^{a \, \mu \nu} \bigg).
    \end{split}
\end{equation}
The BRST invariance of this action is given by the transformations
\be\label{eq3O}
\delta A^a_\mu = D^{ab}_\mu(A) c^b \epsilon 
\ee
\be\label{eq3P}
\delta F^a_{\mu\nu} = g f^{abc} F^b_{\mu\nu} c^c \epsilon 
\ee
\be\label{eq3Q}
\delta \lambda^a_\mu = \left(D^{ab}_{\mu}(A) d^b + g f^{abc} \lambda^b_\mu c^c \right)\epsilon 
\ee
\be\label{eq3R}
\delta \Lambda^a_{\mu\nu} = g f^{abc} \left(F^b_{\mu\nu} d^c + \Lambda^b_{\mu\nu} c^c \right)\epsilon 
\ee
\be\label{eq3S}
\delta {\bar c}^a = -\frac 1 \alpha \partial\cdot A^a \epsilon
\ee
\be\label{eq3T}
\delta {\bar d}^a = -\frac 1 \alpha \partial\cdot\lambda^a \epsilon
\ee
\be\label{eq3U}
\delta c^a = \frac 1 2 f^{abc}  c^b c^c \epsilon 
\ee
\be\label{eq3V}
\delta d^a = f^{abc}  c^b d^c \epsilon .
\ee
In place of Eq. \eqref{eq3N} one could also use
\be\label{eq3W}
\det\left(\begin{array}{cc} 0 & a \\ a & b\end{array}\right)  = -{\det}^2 a 
\ee
which in this case becomes 
\be\label{eq3X}
- {\det}^2 \left(\partial\cdot D^{ab}(A)\right).
\ee
This is the square of the usual Faddeev--Popov determinant that arises
when one use the path integral to quantize ${\cal L}^{(1)}_{\text{YM}}$ of
Eq. \eqref{eq:22}.


The shift of Eq.~\eqref{eq:25} is now modified so that
\begin{equation}\label{eq:673}
    {F}_{\mu \nu}^{a} = {H}_{\mu \nu}^{a} + {f}_{\mu \nu}^{a} + \left ( D^{ab}_{\mu} \lambda^{b}_{\nu} - D_{\nu}^{ab} {\lambda}_{\mu}^{b} \right ) - 2 \Lambda^{a}_{\mu \nu} - 2 J^{a}_{\mu \nu},
\end{equation}
followed by
\begin{equation}\label{eq:674}
    \Lambda^{a}_{\mu \nu} = \Phi^{a}_{\mu \nu} +\frac{1}{2} \left (  D^{ab}_{\mu}(A)\lambda^{b}_{\nu} - D_{\nu}^{ab}(A){\lambda}_{\mu}^{b} + K_{\mu \nu}^{a}  - 2 {J}_{\mu \nu}^{a}\right )
\end{equation}
then Eq.~\eqref{eq:670} becomes
\begin{equation}\label{eq:675}
    \begin{split}
        Z[j,J,k,K] = & \int \mathcal{D} {H}_{\mu \nu}^{a} \mathcal{D} {A}_{\mu}^{a} \int \mathcal{D} {\Phi}_{\mu \nu}^{a} \mathcal{D} {\lambda}_{ \mu}^{a} 
    \int \mathcal{D} \bar{c}^{a} \mathcal{D} c^{a} \int \mathcal{D }
    \bar{d}^{a} \mathcal{D} d^{a} \exp i \int \mathop{d^{d} x}\bigg( 
        \frac{1}{4} {H}_{\mu \nu}^{a} {H}_{}^{a \, \mu \nu} -  {\Phi}_{\mu \nu}^{a} {\Phi}_{}^{a \, \mu \nu} - \frac{1}{4} {f}_{\mu \nu}^{a} {f}_{}^{a \, \mu \nu}\\
                     & - \frac{1}{2} (f^{a \, \mu \nu} - K^{a \, \mu \nu} )(D^{ab}_{\mu} \lambda^{b}_{\nu} - D_{\nu}^{ab} {\lambda}_{\mu}^{b}  - 2 J^{a}_{\mu \nu}) + \frac{1}{4} K_{\mu \nu}^{a} K^{a \, \mu \nu} - \frac{1}{2 \alpha} (\partial^{\mu} A^{a}_{\mu} )^2 -\frac{1}{\alpha} \partial^{\mu} A^{a}_{\mu} \partial^{\nu} \lambda^{a}_{\nu} \\ & + \bar{c}^{a} \partial^{\mu} D^{ab}_{\mu} (A + \lambda ) c^{b} + \bar{c}^{a} \partial^{\mu} D^{ab}_{\mu} (A ) d^{b} + \bar{d}^{a} \partial^{\mu} D^{ab}_{\mu} (A) c^{b} + {j}_{\mu}^{a} A^{a \, \mu} +  k_{\mu}^{a} \lambda^{a \, \mu}  \bigg).
    \end{split}
\end{equation}
In Eq.~\eqref{eq:675}, the integrals over $ {H}_{\mu \nu}^{a} $ and $ {\Phi}_{\mu \nu}^{a} $ now just contribute with an overall constant to $Z$ that can be discarded. Unlike Eq.~\eqref{eq:26}, $J_{\mu \nu}^{a} $ does not enter Eq.~\eqref{eq:675} quadratically, although there are terms proportional to $K^2$ and $JK$, which is a consequence of the tree level propagators for $ {H}_{\mu \nu}^{a} $ and $ {\Phi}_{\mu \nu}^{a} $. If $ {J}_{\mu \nu}^{a} = {K}_{\mu \nu}^{a} = 0$ then Eq.~\eqref{eq:675} reduces to the generating functional for second order YM theory when it is supplemented by a LM field $ \lambda_{\mu \nu}^{a}$.

Eq.~\eqref{eq:675} can be more directly obtained by first integrating over the LM field $ \Lambda^{a \, \mu \nu} $ in Eq.~\eqref{eq:670}; the resulting $ \delta$-function
\begin{equation}\label{eq:61sm}
    \mathop{\delta} \left ( F^{a \, \mu \nu} - f^{a \, \mu \nu} + K^{a \, \mu \nu} \right ),
\end{equation}
means that $ F^{a \, \mu \nu} $ can be replaced by $f^{a\, \mu \nu}-K^{a \, \mu \nu} $. Finally integrating over $F^{a \, \mu \nu} $ Eq.~\eqref{eq:670} becomes
\begin{equation}
    \label{eq:62sm}
\begin{split}
        Z[j,J,k,K] = & \int  \mathcal{D} {A}_{\mu}^{a} \int \mathcal{D} {\lambda}_{ \mu}^{a} 
    \int \mathcal{D} \bar{c}^{a} \mathcal{D} c^{a} \int \mathcal{D }
        \bar{d}^{a} \mathcal{D} d^{a} \exp i \int \mathop{d^{d} x} \bigg( - \frac{1}{4} {f}_{\mu \nu}^{a} {f}_{}^{a \, \mu \nu}  + \lambda^{a}_{\nu} D^{ab}_{\mu} f^{b \,\mu \nu } - \lambda^{a}_{\nu}  D_{\mu}^{ab} {K}^{b \, \mu \nu } \\ & - \frac{1}{2 \alpha} (\partial^{\mu} A^{a}_{\mu} )^2  -\frac{1}{\alpha} \partial^{\mu} A^{a}_{\mu} \partial^{\nu} \lambda^{a}_{\nu} + \bar{c}^{a} \partial^{\mu} D^{ab}_{\mu} (A + \lambda ) c^{b} + \bar{c}^{a} \partial^{\mu} D^{ab}_{\mu} (A ) d^{b} + \bar{d}^{a} \partial^{\mu} D^{ab}_{\mu} (A) c^{b} \\ & + {j}_{\mu}^{a} A^{a \, \mu} +  k_{\mu}^{a} \lambda^{a \, \mu} + J^{a}_{\mu \nu} f^{a \, \mu \nu}-J^{a}_{\mu \nu} K^{a \, \mu \nu} + \frac{1}{4} K_{\mu \nu}^{a} K^{a \, \mu \nu}     \bigg),
    \end{split}
\end{equation}
which seems slightly different than Eq.~\eqref{eq:675}. However,
integrating $D_{\mu}^{ab} \lambda^{b}_{\nu} - D^{ab}_{\nu} \lambda^{b}_{\mu} $ by parts in Eq.~\eqref{eq:675} we see, apart from inconsequential constant factors, that Eq.~\eqref{eq:675} is equal to Eq.~\eqref{eq:62sm}.
If $K=0$, Eq.~\eqref{eq:62sm} reduces to the generating functional of the second form of the YM theory with a LM field $ \lambda^{a}_{\mu \nu} $ with the extra source term $J^{a}_{\mu \nu} f^{a \, \mu \nu} $ like we have in Eq.~\eqref{eq:26}. Thus Eq.~\eqref{eq:62sm} may be seen as the analogous of Eq.~\eqref{eq:26} for the YM theory supplemented with LM fields $ \lambda^{a}_{\mu} $ and $ \Lambda^{a}_{\mu \nu} $ and their corresponding sources given by $ k^{a \, \mu} $ and $ K^{a \, \mu \nu}$ respectively.

\section{First Order Einstein-Hilbert theory with Lagrange Multipliers}

We now will use $Z$ in Eq.~\eqref{eq:469} to obtain the generating function for the first order EH Lagrangian when it is supplemented by LM fields $ \lambda_{\mu \nu} $ and $ \Lambda^{\lambda}_{\mu \nu} $ to ensure that the equation of motion for $
\phi^{\mu \nu} $ and $ G^{\lambda}_{\mu \nu} $ are satisfied. 
The same procedure that was used with the first order YM theory
supplemented by LM fields can be employed here.
From Eq.~\eqref{eq:326} we have that
\begin{equation}\label{eq:776}
    \begin{split}
Z[j,J,k,K] & =   \int \mathcal{D} {G}_{\mu \nu}^{\lambda} \mathcal{D}
        {\phi}_{}^{\mu \nu} \int \mathcal{D} {\Lambda}_{\mu \nu}^{\lambda}
        \mathcal{D} {\lambda}_{}^{\mu \nu} \exp  \biggm\{ i \int d^{d}x
            \biggm[ \frac{1}{\kappa^{2} } \biggm(- \kappa {G}_{\mu \nu}^{\lambda} \phi^{\mu \nu}_{, \lambda} \\
            &+
\frac{1}{2} {G}_{\mu \nu}^{\lambda} [m + \kappa M(\phi)]
       \mbox{}_{\lambda}^{\mu \nu} \mbox{}_{\sigma}^{\pi \tau} {G}_{\pi \tau}^{\sigma} 
            \\ & +
            {\Lambda}_{\mu \nu}^{\lambda} \left (- \kappa \phi^{\mu
            \nu}_{, \lambda} + M\mbox{}_{\lambda}^{\mu \nu} \mbox{}_{\sigma}^{\pi \tau}
            (h) {G}_{\pi \tau}^{\sigma}  \right ) + \lambda^{\mu \nu} \left (\kappa
        {G}_{\mu \nu , \lambda}^{\lambda} \right ) + \frac{1}{2} G^{\lambda}_{\mu \nu}
        \left [ \kappa M\mbox{}_{\lambda}^{\mu \nu} \mbox{}_{\sigma}^{\pi \tau} (
    \lambda ) \right ] {G}_{\pi \tau}^{\sigma} \biggm)  
            \\ & +
            {\mathcal{L} }_{\text{gf}}^{}(\phi , \lambda ) 
            +  j_{\mu \nu} \phi^{\mu \nu} + k_{\mu \nu} \lambda^{\mu \nu} + J^{\mu \nu}_{\lambda} G_{\mu \nu}^{\lambda} +  K^{\mu \nu}_{\lambda} 
    {\Lambda}_{\mu \nu}^{\lambda} \biggm] \biggm\}{\Delta}_{\text{FP}}^{} (\phi, \lambda ).
    \end{split} 
\end{equation}
If we work in the de Donder gauge $\partial_{\mu} \phi^{\mu \nu} = 0$, then
\begin{equation}\label{eq:777}
    {\mathcal{L}}_{\text{gf}}^{} ( \phi, \lambda) = - \frac{1}{2 \alpha} \left [
            (\partial_{\mu} \phi^{\mu \nu} )^{2} + 2 (\partial_{\mu}
            \phi^{\mu \nu} ) \eta_{\nu \sigma} ( \partial_{\beta}
            \lambda^{\beta \sigma} )
        \right ].
\end{equation}

In Eq.~\eqref{eq:776}, integrating over $ \Lambda_{\mu \nu}^{\lambda} $, we get a
\begin{equation}\label{eq:71sm}
\delta \left ( M \mbox{}_{\lambda}^{\mu \nu} \mbox{}_{\sigma}^{\pi \tau}(h)G^{\sigma}_{\pi \tau} - \kappa  \phi_{, \lambda}^{\mu \nu } + \kappa^{2}  K^{\mu \nu}_{\lambda }\right ) \equiv \delta \left ( M \mbox{}_{\lambda}^{\mu \nu} \mbox{}_{\sigma}^{\pi \tau}(h)( G - \tilde{G}[h, K])^{\sigma}_{\pi \tau } \right ),
\end{equation}
where 
\begin{equation} \label{eq:72asm}
\tilde{G}^{\sigma}_{\pi \tau }[h, K] = (M^{-1}) \mbox{}_{\pi \tau}^{\sigma}\mbox{}_{\alpha \beta}^{\rho}(h) \left(h_{, \rho}^{\alpha \beta} - \kappa^{2}  K^{\alpha \beta}_{\rho}\right).
\end{equation}
(Note that $ \tilde{G}_{\pi \tau}^{\sigma} [h,0]$ is the classical value of $G^{\sigma}_{\pi \tau } $.) Integrating now over $ G_{\mu \nu}^{\lambda} $, apart from the determinant $ \det M^{-1}(h)$ that is not relevant\footnote{This determinant, like that of Eq.~\eqref{eq:329}, do not contribute to $Z$ since in a diagrammatic expansion of this the propagators that occur are simply $ {(m^{-1})}{}_{\mu \nu}^{\lambda} {}_{\pi \tau}^{\sigma}$, leaving only tadpole integrals that vanish when using dimensional regularization \cite{Brandt:2016eaj}.}, gives 
\begin{equation}\label{eq:72sm}
    \begin{split}
Z[j,J,k,K] & =    
        \int \mathcal{D} {\phi}^{}_{\mu \nu}\mathcal{D} {\lambda}^{}_{\mu \nu} \exp  \biggm[ i \int d^{d}x \biggm( \frac{1}{2 \kappa^{2}}\tilde{{G} }^{\lambda}_{\mu \nu}[h, K] M\mbox{}^{\mu \nu}_{\lambda} \mbox{}^{\pi \tau}_{\sigma}( \chi ) (  \tilde{{G} } [h, K]- 2 \tilde{G}  [\chi , J] )\mbox{}^{\sigma}_{\pi \tau }
        \\ & +
            {\mathcal{L} }_{\text{gf}}^{}(\phi , \lambda ) 
            +  j_{\mu \nu} \phi^{\mu \nu} + k_{\mu \nu} \lambda^{\mu \nu}  \biggm) \biggm] {\Delta}_{\text{FP}}^{} (\phi, \lambda ),
    \end{split} 
\end{equation}
where $ {\chi}_{}^{\mu \nu} \equiv \eta^{\mu \nu} + \kappa ( \phi^{\mu \nu} + \lambda^{\mu \nu} ) = h^{\mu \nu} + \kappa \lambda^{\mu \nu} $.

The source $J^{\mu \nu}_{\lambda} $ enters Eq.~\eqref{eq:72sm} in a very non-trivial way, interacting with $ \phi^{\mu \nu} $ and $ \lambda^{\mu \nu} $ in a complicated manner. However, if $ \lambda^{\mu \nu} = 0$ in Eq.~\eqref{eq:72sm}, then we simply have 
\begin{equation}\label{eq:783}
\begin{split}
    & \tilde{{G} }^{\lambda}_{\mu \nu}[h, K] M {}{}^{\mu \nu}_{\lambda} {}^{\pi \tau}_{\sigma}( h )   \tilde{{G} }{}^{\sigma}_{\pi \tau} [h , K]- 2 \tilde{{G} }^{\lambda}_{\mu \nu}[h , K] M {}{}^{\mu \nu}_{\lambda} {}^{\pi \tau}_{\sigma}( h )\tilde{G}{}^{\sigma}_{\pi \tau} [h, J]  
    = - \kappa^{2} {\phi}_{, \lambda}^{\mu \nu} {( M^{-1} )}{}_{\mu \nu}^{\lambda} {}_{\pi \tau}^{\sigma} {\phi}_{, \sigma}^{\pi \tau} \\ & - 2\kappa^{2} {J}_{\lambda}^{\mu \nu} {( M^{-1} )}{}_{\mu \nu}^{\lambda} {}_{\pi \tau}^{\sigma}(h) {\phi }_{,\sigma}^{\pi \tau} + \kappa^{4} {K}_{\lambda}^{\mu \nu} {( M^{-1} )}{}_{\mu \nu}^{\lambda} {}_{\pi \tau}^{\sigma}(h) {K}_{\sigma}^{\pi \tau} - 2 \kappa^{4} {K}_{\lambda}^{\mu \nu} {( M^{-1} )}{}_{\mu \nu}^{\lambda} {}_{\pi \tau}^{\sigma}(h) {J}_{\sigma}^{\pi \tau}.
\end{split}
\end{equation}
If $J=K=0$, then Eq.~\eqref{eq:72sm} reduces to the generating function that follows from the second order EH action. From Eq.~\eqref{eq:783}, we see that the dependence on $ J^{\mu \nu}_{\lambda} $ and $ K^{\mu \nu}_{\lambda} $ in Eq.~\eqref{eq:72sm} is much like the dependence of $ Z[j,J,k,K]$ in Eq.~\eqref{eq:62sm} on $ J^{a}_{\mu \nu} $, $ K^{a}_{\mu \nu} $, there being terms proportional to $ K^2$ and $JK$, but not $J^2$.

We could also make the shift
\begin{equation}\label{eq:778}
{G}_{\mu \nu}^{\lambda} \to G_{\mu \nu}^{\lambda} - (M^{-1}){}_{\mu \nu}^{\lambda}\mbox{}_{\pi \tau}^{\sigma}( \chi )  \left( - {\chi}_{, \sigma}^{\pi \tau} + {M}\mbox{}_{\sigma}^{\pi \tau}\mbox{}_{\rho}^{\alpha \beta} (h){\Lambda}_{\alpha \beta}^{\rho} + \kappa^{2} {J}_{\sigma}^{\pi \tau} \right)
\end{equation}
in Eq.~\eqref{eq:776}, followed by 
\begin{equation}\label{eq:779}
    \Lambda_{\mu \nu}^{\lambda} \to {\Lambda}_{\mu \nu}^{\lambda} + (\Xi^{-1}){}_{\mu \nu}^{\lambda} {}_{\pi \tau}^{\sigma} \zeta^{\pi \tau}_{\sigma},
\end{equation}
where
\begin{equation}\label{eq:781}
    {\Xi}{}_{\lambda}^{\mu \nu} {}_{\sigma}^{\pi \tau} = [M(h) M^{-1}(\chi) M(h)]{}^{\mu \nu}_{\lambda} {}^{\pi \tau}_{\sigma}
\end{equation}
and 
\begin{equation}\label{eq:782}
    \zeta^{\mu \nu}_{\lambda} = [M(h) M^{-1}( \chi )] \mbox{}_{\lambda}^{\mu \nu}\mbox{}_{\pi \tau}^{\sigma} ( {\chi}_{, \sigma}^{\pi \tau} - \kappa^{2} {J}_{\sigma}^{\pi \tau} ) - \kappa {\phi}_{, \lambda}^{\mu \nu} + \kappa^{2} K^{\mu \nu}_{\lambda}.
\end{equation}
This results in Eq.~\eqref{eq:776} becoming 
\begin{equation}\label{eq:780}
    \begin{split}
Z[j,J,k,K] = &   \int \mathcal{D} {G}_{\mu \nu}^{\lambda} \exp \left[i \int d^{d}x \left (  \frac{1}{2 \kappa^{2}} {G}_{\mu \nu}^{\lambda} {M}\mbox{}_{\lambda}^{\mu \nu}\mbox{}_{\sigma}^{\pi \tau}( \chi) {G}_{\pi \tau}^{\sigma} \right ) \right] \\
        &\times\int \mathcal{D}  {\Lambda}_{\mu \nu}^{\lambda} \exp \left[i \int d^{d}x \left(-\frac{1}{2 \kappa^{2}} {\Lambda}_{\mu \nu}^{\lambda} \Xi \mbox{}^{\mu \nu}_{\lambda} \mbox{}^{\pi \tau}_{\sigma} {\Lambda}_{\pi \tau}^{\sigma } \right) \right]\\ 
        & \times 
        \int \mathcal{D} {\phi}_{}^{\mu \nu}\mathcal{D}
        {\lambda}_{}^{\mu \nu} \exp  \biggm[ i \int d^{d}x \biggm(
        \frac{1}{2 \kappa^{2}} \left [ {\zeta}_{\lambda}^{\mu \nu}
          (\Xi^{-1}) \mbox{}_{\mu \nu}^{\lambda}\mbox{}_{\pi
            \tau}^{\sigma} {\zeta}_{\sigma}^{\pi \tau} \right.\\
        &\left.- ( \chi^{\mu \nu}_{, \lambda} - \kappa^{2} J^{\mu \nu}_{\lambda}){( M ^{-1})}\mbox{}_{\mu \nu}^{\lambda}\mbox{}_{\pi \tau}^{\sigma}(\chi)( \chi^{\pi \tau}_{, \sigma} - \kappa^{2} {J}_{\sigma}^{\pi \tau} ) \right ] 
            \\ & +
            {\mathcal{L} }_{\text{gf}}^{}(\phi , \lambda ) 
            +  j_{\mu \nu} \phi^{\mu \nu} + k_{\mu \nu} \lambda^{\mu \nu}  \biggm) \biggm] {\Delta}_{\text{FP}}^{} (\phi, \lambda ).
    \end{split} 
\end{equation}
In Eq.~\eqref{eq:780}, the integrals over $ G^{\lambda}_{\mu \nu} $ and $ \Lambda^{\lambda}_{\mu \nu}$ lead to functional determinants $\det^{-1/2} M(\chi) \det^{-1/2} \Xi$ that results in the same determinant $ \det M^{-1}(h)$ obtained above which can be discarded. Rewriting Eq.~\eqref{eq:782} as 
\begin{equation}\label{eq:73sm}
    \zeta^{\mu \nu}_{\lambda} = M {}_{\lambda}^{\mu \nu} {}_{\sigma}^{\pi \tau} (h) ( \tilde{G}[\chi ,J] - \tilde{G} [h, K]  ) {}_{\pi \tau}^{\sigma}
\end{equation}
is easy to see, apart from those irrelevant determinants, that Eq.~\eqref{eq:780} is equal to Eq.~\eqref{eq:72sm}. 

Even though being a more involved approach it is interesting to show
the shifts in Eqs.~\eqref{eq:778} and \eqref{eq:779}. Especially if one rewrites the shift in Eq.~\eqref{eq:778} using Eqs.~\eqref{eq:782} and \eqref{eq:73sm} as
\begin{equation} \label{eq:74sm}
    {G}_{\mu \nu}^{\lambda} \to G_{\mu \nu}^{\lambda} + \tilde{G}^{\lambda}_{\mu \nu} [\chi , J]- (M^{-1}){}_{\mu \nu}^{\lambda} {}_{\pi \tau}^{\sigma}( \chi)  M {}{}^{\pi \tau}_{\sigma} {}^{\rho}_{\alpha \beta}(h){\Lambda}^{\alpha \beta}_{\rho}
\end{equation}
which makes clear the analogy between them and the shifts in Eqs.~\eqref{eq:673} and \eqref{eq:674}. These shifts lead to generating functional of Eq.~\eqref{eq:675} which is consequently  analogous to the generating functional of Eq.~\eqref{eq:780}. In addition, the shifts in Eqs.~\eqref{eq:673} and \eqref{eq:674} are the generalization of the shift in Eq.~\eqref{eq:25} when LM fields are present. Then, as Eq.~\eqref{eq:62sm} is the analogues of Eq.~\eqref{eq:26} for the YM theory with LM fields, so Eq.~\eqref{eq:780} for the EH theory (in this case with the LM fields $ \lambda^{\mu \nu } $, $ \Lambda^{\lambda }_{\mu \nu} $; whose sources are respectively $ k_{\mu \nu } $, $ K^{ \mu \nu}_\lambda $).

\section{Discussion}

We have examined various aspects of quantizing the YM and EH actions in first order form by using the path integral. First of all, it is shown that the first and second order forms of these theories are equivalent. By examining how the generating functional in these two forms depends on the source function for the auxiliary field, we are able to find structural identities that relate Green's functions computed using either the form of the generating functional that follows from the first or second order form of these theories.

Next, the consequences of including a LM field to ensure that the
classical equations of motion are satisfied 
are analyzed in some detail. It has been know that by using such a LM field, radiative effects beyond one-loop order do not contribute to the generating functional \cite{McKeon:1992rq,Brandt:2018lbe, PhysRevD.100.125014}. Here we show how this occurs with various choices for the source functions, and also consider the effects of having a choice of variables that leaves the propagators in these models all diagonal.

Then, the use of LM fields in conjunction with gauge theories is considered, with particular attention paid to the first order form of the YM and EH actions.
Since radiative effects beyond one-loop order all vanish when using the LM field, it is possible in YM theory to find an exact relation between bare and renormalized quantities when using dimensional regularization, so that one can find the renormalization group $ \beta $-function exactly \cite{McKeon:1992rq}. It has been shown that in $n=4 - \epsilon $ dimensions the divergences
arising in the EH action from the usual one-loop diagrams are \cite{tHooft:1974bx, tHooft:2002xp}
\begin{equation}\label{eq:342}
    \Delta \mathcal{L} = \frac{\sqrt{-g}}{8 \pi^{2} \epsilon} \left (
        \frac{1}{120} R g^{\mu \nu} + \frac{7}{20} R^{\mu \nu}
    \right ) R_{\mu \nu}.
\end{equation}
and so by Eqs.~\eqref{eq:776} and ~\eqref{eq:342} this divergence can be absorbed by the shift
\begin{equation}\label{eq:343}
    \lambda^{\mu \nu} \rightarrow \lambda^{\mu \nu} - \frac{\sqrt{-g}}{4 \pi^{2} \epsilon} \left (
        \frac{1}{120} R g^{\mu \nu} + \frac{7}{20} R^{\mu \nu}
    \right ).
\end{equation}
The last term in Eq.~\eqref{eq:343} is twice the one-loop contribution given by Eq.~\eqref{eq:342}, since when the LM fields $ \lambda^{\mu \nu} $ and $ \Lambda^{\lambda}_{\mu \nu}$ are employed the one-loop contributions are doubled \cite{PhysRevD.100.125014}. No higher loop divergences like those of ref.\ \cite{Goroff:1986th, vandeVen:1992gw, PhysRevD.100.026018} arise as well.
The BRST invariance of Eq.~\eqref{eq:776} ensures that the theory is unitary \cite{PhysRevD.100.125014}.

It would be of interest when considering the EH action to consider source functions that are used in conjunction with the curvature tensor $ R_{\mu \nu} $; in Eq.~\eqref{eq:776} this would involve inserting a term of the form 
\begin{equation}\label{eq:335}
    \mathfrak{J}^{\mu \nu} R_{\mu \nu} (G) \equiv \mathfrak{J}^{\mu \nu} \left( {G}_{ \mu \nu, \lambda}^{ \lambda} + \frac{1}{d-1} {G}_{ \mu \lambda}^
{\lambda} {G}_{ \nu \sigma}^{ \sigma} - {G}_{ \mu \sigma}^{ \lambda}
{G}_{ \nu \lambda}^{ \sigma}\right)
\end{equation}
into the action.

One problem that needs to be addressed is to examine the renormalizability of the EH action when the classical equations of motion are imposed on $ h^{\mu \nu} $ and $ {G}_{\mu \nu}^{\lambda} $, but these fields are also coupled to matter fields by whose equations of motion are not so restricted. One may then expect the radiative corrections to vanish beyond one-loop order, if LM fields are also used with matter fields to restrict their configurations to those that satisfy the equations of motion. We currently are considering this issue.

\begin{acknowledgments}
{Roger Macleod had a useful suggestion. 
F. T. B. and J. F. thank CNPq (Brazil) for financial
support. S. M.-F. thanks CAPES (Brazil) for financial support.
This study was financed in part by the Coordena\c{c}\~{a}o de Aperfei\c{c}oamento 
de Pessoal de N\'{\i}vel Superior - Brasil (CAPES) - Finance Code 001. This work comes as an aftermath of an original 
project developed with
the support of FAPESP (Brazil),  grant number 2018/01073-5.}
\end{acknowledgments}

\appendix

\section{The diagonal formulation of the Lagrange multiplier theory}
In this appendix we show that one can find a general formulation of the LM theory\footnote{We will use the generic term ``LM theory'' for a theory supplemented with a LM field which ensure that the equations of motion are satisfied.} without mixed propagators which simplifies the computation of amplitudes. It is a diagonal formulation (similar formulations for the first order YM and EH theories are found in ref.\ \cite{Brandt:2016eaj, Brandt:2018avq}) of the general formalism that appears in ref.\ \cite{PhysRevD.100.125014}. The diagonal formulation of the massless scalar field theory with LM fields is shown in section IV.

Consider the Lagrangian $
{\mathcal{L}}_{\text{}}^{} [ \phi_i] $ for a field $ \phi_i$ with the addition of a source term $j_i \phi_i$ (such source term is needed to obtain the Green's functions from the generating functional coming from the path integral quantization of the theory). Its action with a LM field $ \lambda_j $ which
imposes the equations of motion on the field $ \phi_{i} $ is
\begin{equation}\label{eq:C1}
    S[\phi_i,\lambda_i] = \int \mathop{d^{d}x} \left [ \mathcal{L} ( \phi_{i} )  + \lambda_j\left ( \frac{\partial
    \mathcal{L}[ \phi_{i} ]}{\partial \phi_j} + j_i \right ) + j_i \phi_{i}  \right ].
\end{equation}
Hence, the generating functional arising from the path integral quantization procedure is of the form \eqref{eq:447}
\begin{equation}\label{eq:C2}
    Z[j_i, k_i] = \int \mathcal{D} \phi_i \mathcal{D} \lambda_i \exp i \left(S[\phi_{i},\lambda_i] + \int
        \mathop{d^{d}x} k_i
    \lambda_i\right),
\end{equation}
where we also introduce an independent source $k_i$ for the LM field $
\lambda_i$. Doing the shift 
\begin{equation}\label{eq:shift1}
    \phi_i = \varphi_i - \lambda_i,
\end{equation}
we obtain an equivalent generating functional given by
\begin{equation}\label{eq:Zreg3}
    Z'[j_i, k_i] = \int \mathcal{D} \varphi_i \mathcal{D} \lambda_i \exp i
    \left (S[\varphi_i - \lambda_i,\lambda_i]+\int \mathop{d^{d} x}
    k_i \lambda_i \right).
\end{equation}
The action $S' \equiv S[\varphi_i - \lambda_i,\lambda_i]$ can be expanded around the
field $ \varphi_i$ as (treating $ \phi_{i} $ and $ \lambda_{i} $ as Bosonic fields)
\begin{equation}\label{eq:explag1}
    S' = \int \mathop{d^{d}x} \left ( \mathcal{L}( \varphi_i ) +
        \sum_{n=2}^{\infty} (- 1)^{n+1} \frac{n-1}{n!} \lambda_j^{n}
        \frac{\partial^{n} \mathcal{L} (\varphi_i)}{\partial
    \varphi_j^{n} } + j_i \varphi_i  \right).
\end{equation}
This does not contain, at quadratic order, any mixing of the fields $ \phi_i $ and $ \lambda_i$ as were in the action Eq.~\eqref{eq:C1}. Since the generating
functional in Eq.~\eqref{eq:Zreg3} does not lead to mixed propagators, that is,
$ \left \langle \varphi_i\lambda_j \right \rangle =\left \langle \lambda_i \varphi_j \right \rangle = 0$, it can be interpreted as being the generating functional of the diagonal formulation of the LM
formalism. Moreover, by setting $n=2$ at
Eq.~\eqref{eq:explag1} one sees that the bilinear terms in the LM field $\lambda _i$ 
are just equal to the negative of the term bilinear in the field
$\varphi_i$. This implies that their propagators are the negative of each other
\be
\label{eq:C5a}
 \left \langle \varphi_i \varphi_j\right
\rangle=- 
\left \langle \lambda_i \lambda_j \right \rangle = \left \langle \phi_i \phi_j \right \rangle_{\lambda = 0}.
\ee
The propagator of the field $ \varphi_i$ is equal to
the propagator of the field $ \phi_i $ in the form of the theory without the LM field $\lambda_i$; denoted by $ \left \langle \phi_i \phi_j \right \rangle_{\lambda = 0} $. The propagator $ \left \langle \phi_i \phi_j \right \rangle_{\lambda = 0} $ follows directly from the inverse of term bilinear in $ \phi_i$ of the Lagrangian $ \mathcal{L} [ \phi_i ] $. 

Since the generating functional of the LM theory in Eq.~\eqref{eq:C2} is linear in the LM field $
\lambda_i$, it can be integrated yielding a delta function. In the
diagonal formulation of Eq.~\eqref{eq:explag1} non-linear terms appear and we cannot integrate the LM
field in the same way. However, we can relate Green's functions computed using the diagonal
formulation and those obtained using the standard formulation of the LM theory by taking the functional derivative 
of the generating functionals of Eqs.~\eqref{eq:C2} and \eqref{eq:Zreg3} with respect to the sources $j_i$
and $k_i$, as done in \cite{Brandt:2020sho, Brandt:2020vre}. Such relations yield a set of structural identities which are
necessary for the consistency of the diagonal formulation of the LM
formalism.

\subsection{Structural Identities}

To obtain the structural identities we need compare the functional
differentiation of the generating functionals $Z[j_i,k_i]$ and $Z'[j_i,k_i]$ with respect to the
sources $j_i$ and $k_j$. We first note that $Z'[0,k_i]=Z[0,k_i]$ so the Green's
function with only external LM fields $ \lambda $ must be the same in both formulations. 
This means the Green's function $ \langle 0|T \lambda_{i_1}(x_1) \cdots \lambda_{i_n}(x_n) | 0 \rangle
$ is the same either in the standard or in the diagonal formalism of the LM theory.

\subsubsection{A first structural identity}
Applying $ {\delta^{2} }/{\delta k_i(x) \delta j_j(y)}$ to the
generating functionals $Z$ and $Z'$, setting $j_j=k_i=0$, and equating the
results; we find the first structural identity
\begin{equation}\label{eq:C5}
    \left \langle 0|T \lambda_i(x) \phi_j (y)|0\right \rangle + \left \langle 0|T \lambda_i(x)\lambda_j (y)|0\right \rangle = \left
\langle 0|T\lambda_i(x) \varphi_j (y) |0\right \rangle,
\end{equation}
where the left side is computed in the usual LM formulation and the right side
in the diagonal formalism.  
The above identity can be easily generalized as
\begin{equation}\label{eq:C6}
    \left \langle 0|T \lambda_{j_1} (x_1) \cdots
    \lambda_{j_n} (x_n) [\phi_i(y) + \lambda_i(y)]|0\right \rangle = \left
\langle 0|T \lambda_{j_1}(x_1) \cdots \lambda_{j_n}(x_n) \varphi_i(y) |0\right \rangle,
\end{equation}
which may be interpreted as the quantum-mechanical generalization of the
classical equality 
\begin{equation}\label{eq:CE}
    \phi_i = \varphi_{i} - \lambda_i
\end{equation}
defined by the shift of Eq.~\eqref{eq:shift1}. 

\subsubsection{More involved structural identities}
We see that each functional differentiation with respect to $j_i(x_i)$ leads to a
factor of $ \phi_i(x_i) + \lambda_{i}(x_i)$ for $Z$ and $ \varphi_i(x_i)$ for $Z'$. 
So we can actually use Eq.~\eqref{eq:CE} directly in any Green's function and obtain a structural identity much like Eq.~\eqref{eq:C5}. Let us take the two-point Green's function (we drop the arguments of the fields; $ \phi_i(x_i)$ is replaced by $ \phi_i$)
\begin{equation}\label{eq:GR1}
    \left \langle 0| T(\phi_i + \lambda_i)(\phi_j + \lambda_j)|0 \right\rangle
\end{equation}
using Eq.~\eqref{eq:CE} we obtain immediately that 
\begin{equation}\label{eq:2si}
    \left \langle 0| T(\phi_i + \lambda_i)(\phi_j + \lambda_j)
        |0\right \rangle = \left \langle 0| T
    \varphi_i \varphi_j |0 \right \rangle.
\end{equation}
Note that left side of Eq.~\eqref{eq:2si} contributes with at most 4 distinct Green's functions. So increasing the number of external $ \phi $ fields leads to structural identities much more involved than Eq.~\eqref{eq:C5}. For example, the
$n$-point generalization of the structural identity of 
Eq.~\eqref{eq:2si} is 
\begin{equation}\label{eq:nsi}
\left \langle 0|T  (\phi_{j_1} + \lambda_{j_1}) \cdots ( \phi_{j_{n}} + \lambda_{j_{n}} ) |0\right \rangle = \left \langle 0|T \varphi_{j_1} \cdots \varphi_{j_n} |0 \right \rangle.
\end{equation}

Expanding the left side of Eq.~\eqref{eq:2si} we have that
\begin{equation}\label{eq:exp2si}
    \left \langle 0| T\phi_i \phi_j |0 \right \rangle + \left \langle 0|
        T \phi_i  \lambda_j  |0 \right \rangle + \left \langle 0| T\lambda_i 
        \phi_j  |0 \right
        \rangle + \left \langle 0|  T\lambda_i 
        \lambda_j |0 \right
        \rangle = \left \langle 0| T\varphi_i \varphi_j |0 \right \rangle.
\end{equation}
Using $ \left < \varphi_i \lambda_j \right > = \left \langle \lambda_i \varphi_j \right \rangle =0 $ in Eq.~\eqref{eq:C5} and Eq.~\eqref{eq:exp2si} at tree level, we have that the following identities hold for the propagators 
\begin{subequations}
\label{eq:FR}
\begin{equation}
    \left \langle  \phi_i \lambda_j  \right \rangle = \left \langle
        \lambda_i \phi_j \right \rangle =  - \left \langle
\lambda_i \lambda_j \right \rangle, \end{equation}
\begin{equation}
 \left
    \langle \phi_i \phi_j \right \rangle= \left \langle \lambda_i
    \lambda_j\right
    \rangle+ \left \langle \varphi_i \varphi_j \right \rangle,
\end{equation}\end{subequations}
With Eq.~\eqref{eq:C5a} and the fact that the LM
propagator $ \left \langle \lambda_i \lambda_j \right \rangle$ is the
same in both formulations, Eqs.~\eqref{eq:FR} yield
\begin{subequations}
\label{eq:CFR}
\begin{equation}
    \left \langle  \phi_i \lambda_j  \right \rangle= \left \langle
        \lambda_i \phi_j \right \rangle =  - \left \langle
    \lambda_i \lambda_j \right \rangle= \left \langle \varphi_i \varphi_j \right \rangle,
    \end{equation}
    \begin{equation}
    \left
    \langle \phi_i \phi_j \right \rangle = 0.
\end{equation}
\end{subequations}
Eq.~\eqref{eq:CFR} agrees with the propagators derived by the inverse of the bilinears terms
of Eqs.~\eqref{eq:C1} and \eqref{eq:Zreg3}.

In the next appendix we will verify some of these identities up to one-loop order
using the $ A^3 $ scalar field theory. As we saw in section 4 such structural identities and the fact that the LM theory restricts the radiative effects to one-loop order imply that the sum of higher loop contributions for any Green's function in the diagonal formulation must vanish. So as an example, we also will check that at two loop-order the sum of the contributions for a two-point Green's function in the diagonal formulation is zero.

\section{Explicit verification for the $ A^{3} $ scalar theory}

Here we will explicitly verify the structural identities Eqs.~\eqref{eq:C5} and \eqref{eq:exp2si} derived in the last section in the simple $ A^3$ scalar theory. First we obtain the Feynman rules for the original theory (in our case the $ A^3$ scalar theory), the original theory with a LM field (LM theory) and the diagonal formulation of the LM theory. In this way, we can compute the Green's functions up to one-loop order and verify explicitly these structural identities.
This simple model will be enough to show how these identities can be verified and used in a practical way. This rather simple model share many similarities to the YM theory, without such complications as color and field spin indices.

Let us start with the Lagrangian 
\begin{equation}\label{eq:C17}
    {\mathcal{L}}_{\text{}}^{(3)}( A )= \frac{1}{2} (\partial_{\mu} A)^2 - \frac{1}{2} m^{2} A^2 - \frac{g}{3!} A^3.
\end{equation}
The introduction of a LM field $ B $ leads to the action 
\begin{equation}\label{eq:C18}
    S^{(3)} [A] = \int \mathop{d^{d}x} \left [ \mathcal{L}^{(3)} ( A)  + B \left ( -\partial_{\mu} \partial^{\mu} A- m^2 A  + j - \frac{g}{2!} A^{2} \right )  + j A  \right ].
\end{equation} 
Using the path integral quantization we define the generating functional  
\begin{equation}\label{eq:C18a}
    Z[j, k] = \int \mathcal{D} C \mathcal{D} B \exp i
    \int \mathop{d^{d} x} \left [ \mathcal{L}^{(3)} ( A)  + B \left ( -\partial_{\mu} \partial^{\mu} A- m^2 A +j - \frac{g}{2!} A^{2} \right )  + j A  + k B \right].
\end{equation}
It leads to the propagators (see Eqs.~\eqref{eq:CFR}) $ \left \langle A A \right \rangle= 0$ and $ \left \langle B A \right \rangle= \left \langle A B \right \rangle=-\left \langle B B  \right \rangle= \left \langle A A \right \rangle_{B =0} $, where $ \left \langle A A \right \rangle_{B =0} $ is the usual massive scalar propagator and the vertices $( A A A )$ and $( A A B )$ both equivalent to the cubic vertex of original theory without LM fields $ (A A A )_{B =0}$. The Feynman rules obtained from the original Lagrangian Eq.~\eqref{eq:C17} are
\begin{subequations}
\begin{equation}
    \vcenter{\hbox{\includegraphics[scale=0.5]{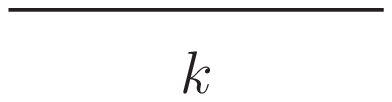}}} \frac{i}{k^2 - m^{2} + i \epsilon},
    \label{fig:FR1a}
\end{equation}
\begin{equation}
    \vcenter{\hbox{\includegraphics[scale=0.5]{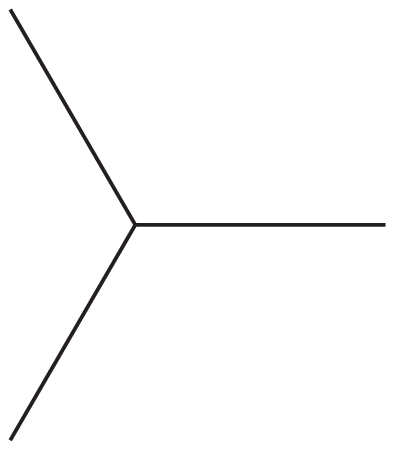}}} -ig,
    \label{fig:FR1b}
\end{equation}
\end{subequations}
where solid lines represent the quantum of the scalar field $ A $ and for the LM theory then are obtained from Eq.~\eqref{eq:C18a}: 
\begin{subequations}
    \label{eq:FRnd}
\begin{equation}
    \vcenter{\hbox{\includegraphics[scale=0.5]{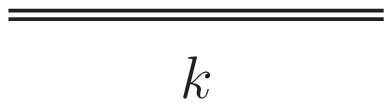}}} \frac{-i}{k^2 - m^{2} + i \epsilon},
    \label{fig:FR2b}
\end{equation}
\begin{equation}
    \vcenter{\hbox{\includegraphics[scale=0.5]{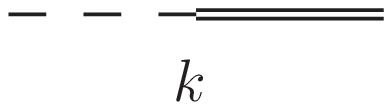}}} \frac{i}{k^2 - m^{2} + i \epsilon},
    \label{fig:FR2c}
\end{equation}
\begin{equation}
    \vcenter{\hbox{\includegraphics[scale=0.5]{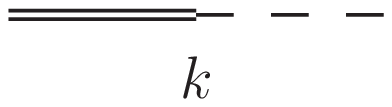}}} \frac{i}{k^2 - m^{2} + i \epsilon},
    \label{fig:FR2c1}
\end{equation}
\begin{equation}
    \vcenter{\hbox{\includegraphics[scale=0.5]{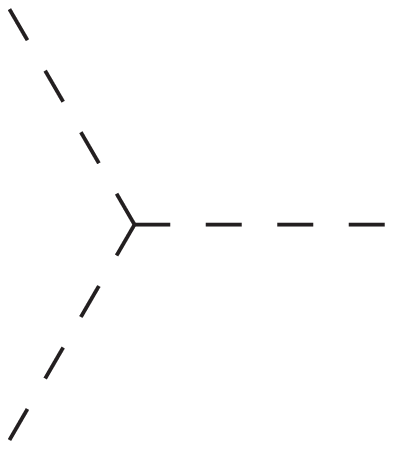}}} -ig,
    \label{fig:FR2d}
\end{equation}
\begin{equation}
    \vcenter{\hbox{\includegraphics[scale=0.5]{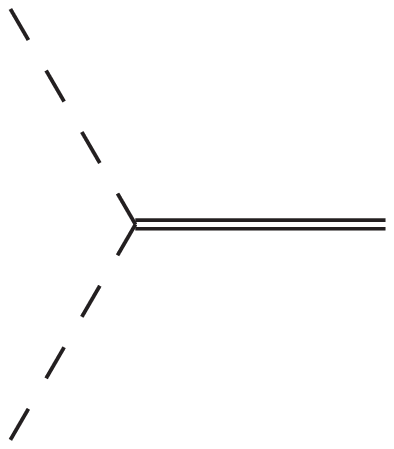}}} -ig,
    \label{fig:FR2e}
\end{equation}
\end{subequations}
where the quanta of the fields $ A $ (LM theory) and the LM field $ B $ are represented respectively by dashed lines and by double solid lines.

For the diagonal formulation the Feynman rules are obtained from Eqs.~\eqref{eq:Zreg3} and \eqref{eq:explag1}. Substituting Eq.~\eqref{eq:C17} into Eq.~\eqref{eq:explag1} yields
\begin{equation}\label{eq:C19}
    Z'[j, k] = \int \mathcal{D} C \mathcal{D} B \exp i
    \int \mathop{d^{d} x} \left ( \mathcal{L}^{(3)} ( C) 
    - \frac{1}{2} (\partial_{\mu} B )^2 + \frac{1}{2} m^{2} B ^2 - \frac{g}{3!} A^3
        - 2g \frac{B^{3}}{3!} + g \frac{B^2}{2!} A + jC + k B \right).
\end{equation}
From Eq. \eqref{eq:C19} we obtain the propagators $ \left \langle C C \right \rangle= - \left \langle B B \right \rangle = \left \langle A A \right \rangle_{ B =0}$ and mixed propagators do not appear. In this formulation, there are now three vertices $ (A A A )$, $ (B B B)$ and $ (B B A)$ that are proportional to the cubic vertex of the original theory. The Feynman rules are 
\begin{subequations}
    \label{eq:FRd}
\begin{equation}
    \vcenter{\hbox{\includegraphics[scale=0.5]{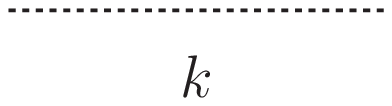}}} \frac{i}{k^2 - m^{2} + i \epsilon},
    \label{fig:FR3a}
\end{equation}
\begin{equation}
    \vcenter{\hbox{\includegraphics[scale=0.5]{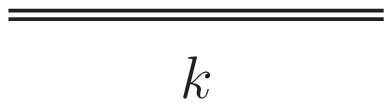}}} \frac{-i}{k^2 - m^{2} + i \epsilon},
    \label{fig:FR3b}
\end{equation}
\begin{equation}
    \vcenter{\hbox{\includegraphics[scale=0.5]{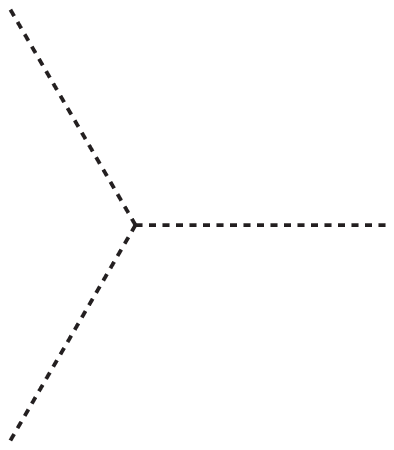}}} -ig, 
    \label{fig:FR3c}
\end{equation}
\begin{equation}
    \vcenter{\hbox{\includegraphics[scale=0.5]{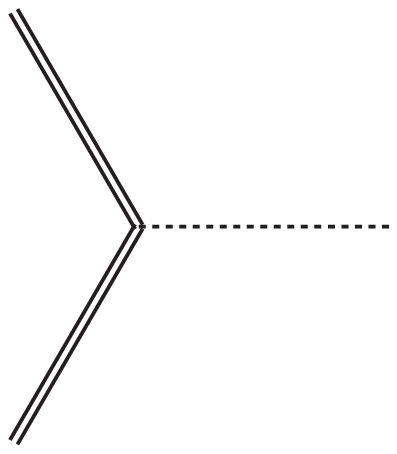}}} ig,
    \label{fig:FR3d}
\end{equation}
\begin{equation}
    \vcenter{\hbox{\includegraphics[scale=0.5]{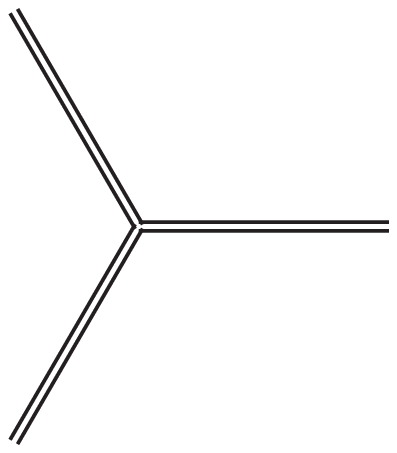}}} -2ig,
    \label{fig:FR3e}
\end{equation}
\end{subequations}
where the quantum of the field $ C $ is represented by dotted lines.

Since the Feynman rules for the theory with the LM field in both formulation, the non diagonal and diagonal formulation, are proportional to the Feynman rules of the original theory, we have that the diagrams obtained from them are also proportional to the analogous diagrams found in the original theory. Thus, we will not need to compute any diagram explicitly, but only compares it relatively to the equivalent diagram in the original theory. In the Fig.\ 7, it is shown some usual perturbative contributions coming from the $ A^3$ scalar theory. So all one-loop diagram that contribute to any two-point Green's function must be proportional to the diagram in the Fig.\ 7(a) and so on.
\begin{figure}[h!]
    \includegraphics[scale=0.37]{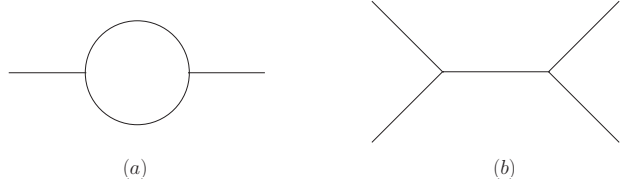}
    \caption{Usual perturbative contributions from the $A^3$ scalar field theory. The diagram (a) is the one-loop contribution for the two-point Green's function. The diagram (b) is the tree level contribution for the 4-point Green's function.}
\end{figure}

\subsection{Verification of Eq.~\eqref{eq:C5}}

Let us verify the structural identity Eq.~\eqref{eq:C5} which in the $ A^3 $ scalar theory reads 
\begin{equation}\label{eq:C20}
    \left \langle 0| T B(x) A(y) |0 \right \rangle + \left \langle 0| T B(x) B(y)|0 \right \rangle = \left \langle 0| T B(x) C(y) |0 \right \rangle.
\end{equation}
By the Feynman rules of Eqs.~\eqref{eq:FRnd} and \eqref{eq:FRd} it is
straightforward to check that, at tree level, Eq.~\eqref{eq:C20} holds. 

At the one-loop order, the first term of Eq.~\eqref{eq:C20} is equal to zero, since there is no propagator for $ A$ in the LM theory. Indeed, any Green's function at one-loop order, with a field $ A $ vanishes, that is,
\begin{equation}\label{eq:Cvan}
    \langle 0|T F_1(x_1) \cdots F_n (x_n) A(y)| 0 \rangle = 0 \quad  \text{(at one-loop order)},
\end{equation}
where $F_i$ may be either a field $ A$ or a LM field $ B$.
In the Fig.\ 8 we have the only diagram that contribute at one-loop order for the second term of Eq.~\eqref{eq:C20}.
Denoting the usual contribution at one-loop order of the $A^{3}$ scalar field theory in Fig.\ 7(a) by $ \langle 0|T A(x) A(y) | 0 \rangle_{B =0}$, we have that
\begin{equation}\label{eq:C20a}
    \langle 0|T B(x) B(y) | 0 \rangle = 2 \langle 0|T A(x) A(y)| 0 \rangle_{B =0}.
\end{equation}
The diagram that contribute for the right side of Eq.~\eqref{eq:C20} at one-loop order is shown in the Fig.\ 9. By the Feynman rules in Eq.~\eqref{eq:FRd}, we see that it also contribute with twice of the usual contribution of the $A^3$ scalar field theory. So that we verified that Eq.~\eqref{eq:C20} holds up to one-loop order. 
\begin{figure}
    \includegraphics[scale=0.27]{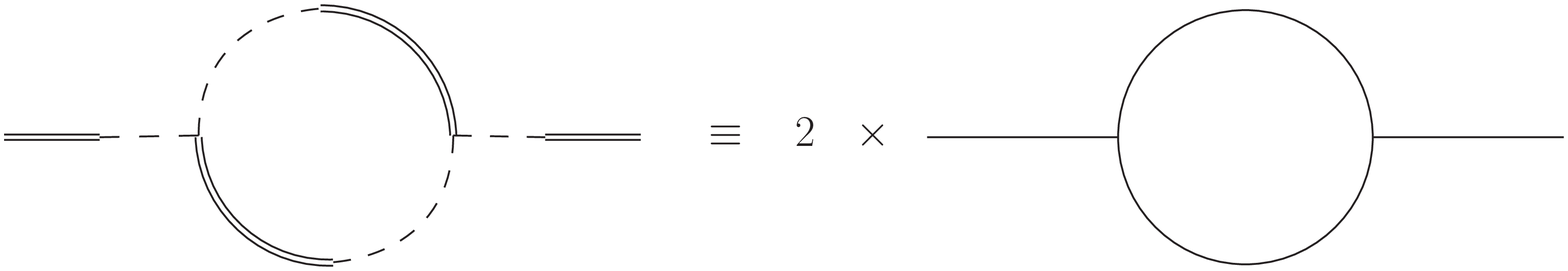}
    \label{fig:A6}
    \caption{The one-loop contribution for $\left \langle 0| T B(x) B(y)|0 \right \rangle$ in the non diagonal formulation of the LM theory. It is equivalent to twice the one-loop contribution of the $A^3$ scalar field theory in the Fig.\ 7(a). }
\end{figure}
\begin{figure}
    \includegraphics[scale=0.27]{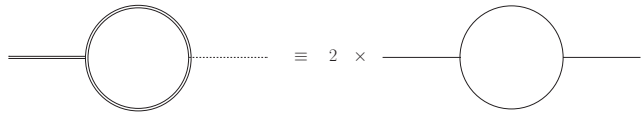}
    \label{fig:A6b}
    \caption{The one-loop contributions for $\left \langle 0| T B(x) C(y) |0 \right \rangle$ in the diagonal formulation of the LM theory. It is equivalent to twice the one-loop contribution of the $A^3$ scalar field theory in the Fig.\ 7(a).}
\end{figure}

Beyond one-loop the left side of Eq.~\eqref{eq:C20} vanishes, since the LM theory restricts the contributions to one-loop order \cite{PhysRevD.100.125014,McKeon:1992rq, Brandt:2018lbe}. Thus the Eq.~\eqref{eq:C20} is simplified to
\begin{equation}\label{eq:C21}
    \left \langle 0| T B(x) C(y) |0 \right \rangle  = 0 \quad \text{(beyond one-loop order).} 
\end{equation}
The two-loop contributions for $ \langle 0|T B(x) C(y) | 0 \rangle$ are shown in the Fig.\ 10. As we can see in Fig.\ 10 each row of contributions cancels out. Then we checked that Eq.~\eqref{eq:C21} holds at two-loop order. Consequently we verified that Eq.~\eqref{eq:C20} is valid up to two-loop order.
\begin{figure}
    \includegraphics[scale=0.5]{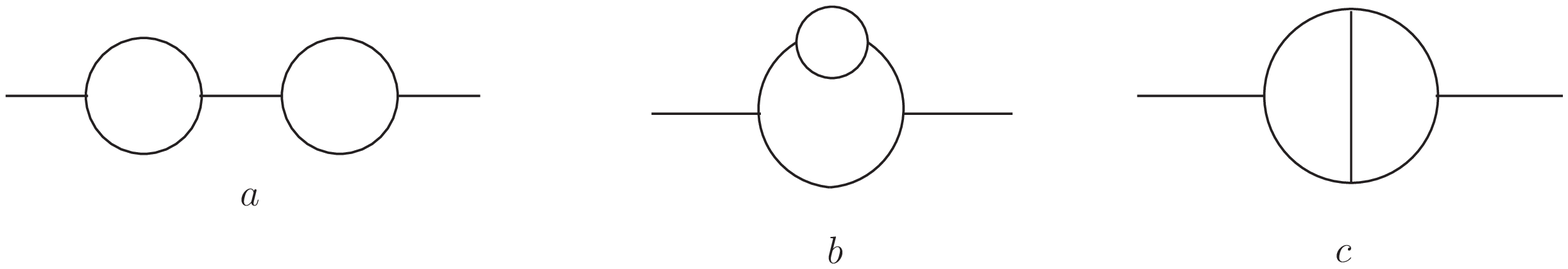}
    \includegraphics[width= 1\textwidth]{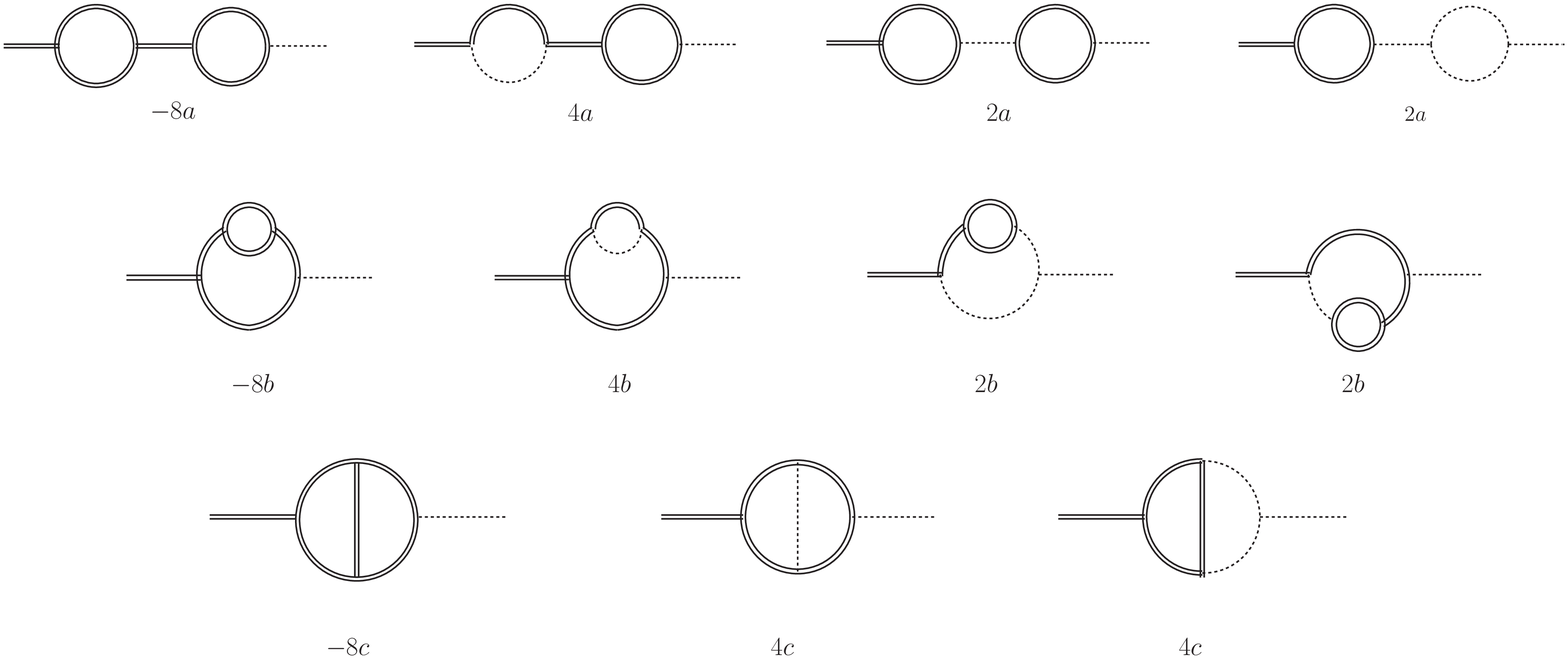}
    \caption{Two-loop contributions for $ \langle 0|T B(x) C(y) | 0 \rangle$ in the diagonal formalism compared with the usual perturbative contributions from the $A^3$ scalar field theory; diagrams $a$, $b$ and $c$.}
\end{figure}

\subsection{Verification of Eq.~\eqref{eq:exp2si}}

The structural identity Eq.~\eqref{eq:exp2si} can be written as 
\begin{equation}\label{eq:C22}
    \left \langle 0| T A(x)  A(y)  |0 \right \rangle + \left \langle 0| T A(x)  B(y)  |0 \right \rangle + \left \langle 0| T B(x) A(y)  |0 \right
        \rangle + \left \langle 0|T B(x) B(y) |0 \right
        \rangle = \left \langle 0| T C(x)  C(y)  |0 \right \rangle.
\end{equation}
With the Feynman rules in Eqs.~\eqref{eq:FRnd} and \eqref{eq:FRd} we easily check that at tree level Eq.~\eqref{eq:C22} holds. To check Eq.~\eqref{eq:C22} at one-loop order we can first use the result Eq.~\eqref{eq:Cvan}. Thus, the identity in Eq.~\eqref{eq:C22} simplifies to
\begin{equation}\label{eq:C23}
    \langle 0|T B(x) B(y) | 0 \rangle = \langle 0|T C(x) C(y) | 0 \rangle \quad \text{(at one-loop order)}.
\end{equation}

The left side of Eq.~\eqref{eq:C23} at one-loop order was already computed. However, let us show that the same result can be found in the diagonal formulation. The one-loop diagrams that contribute are shown in the Fig.\ 11. Using the Feynman rules in Eq.~\eqref{eq:FRd} we have that the first diagram in Fig.\ 11 contributes with $4 \langle 0|T A(x) A(y) | 0 \rangle_{B=0} $ and the second with $-2 \langle 0|T A(x) A(y) | 0 \rangle_{B=0} $. The sum of these contributions results in the same result found in the non diagonal formulation given by Eq.~\eqref{eq:C20a}.
\begin{figure}
    \includegraphics[scale=0.3]{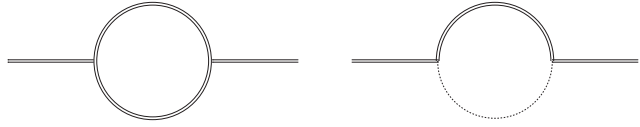}
    \label{fig:A7ex}
    \caption{The one-loop contributions for $\left \langle 0| T B(x) B(y) |0 \right \rangle$ in the diagonal formulation of the LM theory.}
\end{figure}

For the right side of Eq.~\eqref{eq:C23}, at one-loop order, we have the diagrams in the Fig.\ 12, where each diagram is equivalent to the diagram in Fig. 7(a). Thus, the total contribution at one-loop order is 
\begin{equation}\label{eq:C42}
    \langle 0|T C(x) C(y) | 0 \rangle = 2 \langle 0|T A(x) A(y) | 0 \rangle_{B = 0},
\end{equation}
which is equal to Eq.~\eqref{eq:C20a}. So we have checked that Eq.~\eqref{eq:C23} is valid up to one-loop order, and consequently Eq.~\eqref{eq:C22} as well.
\begin{figure}[ht!]
    \includegraphics[scale=0.3]{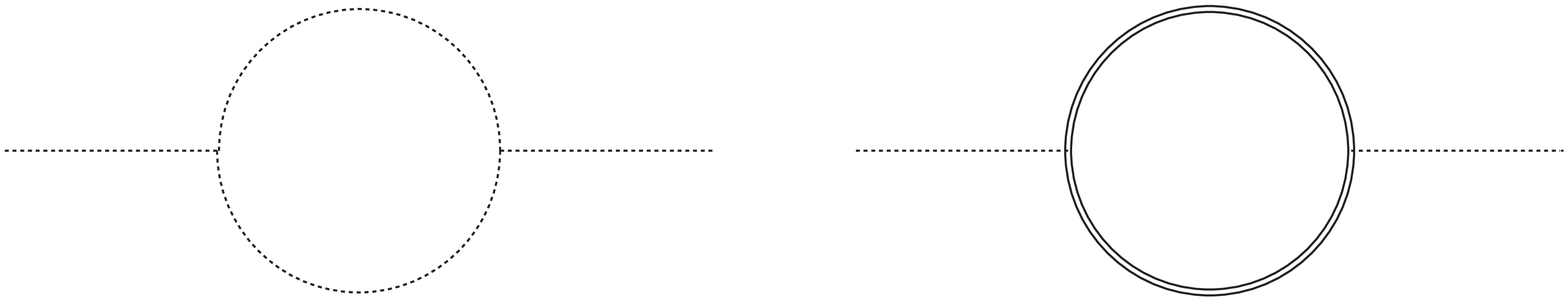}
    \label{fig:A7}
    \caption{The one-loop contributions for $\left \langle 0| T C(x) C(y)|0 \right \rangle$ in the diagonal formulation of the LM theory.}
\end{figure}

\end{document}